\documentclass[aps,
               prd,
               twocolumn,
               superscriptaddress,
               amsfonts,
               amssymb,
               amsmath,
               preprintnumbers,
               nofootinbib,
               unsortedaddress,
               showpacs,
               tightenlines,
               floats,
               letterpaper,
               eqsecnum,
               ]{revtex4}

\usepackage{graphicx}
\usepackage{latexsym}
\usepackage{bm}
\usepackage[draft=false]{hyperref}

\def\F{{\mathcal F}}

\begin{document}

\title{
Constraints on the mass spectrum of primordial black holes and braneworld parameters from the high-energy diffuse photon background
}

\author{Yuuiti Sendouda}\email[]{sendouda@utap.phys.s.u-tokyo.ac.jp}
\affiliation{Department of Physics, School of Science, University of Tokyo, 7-3-1 Hongo, Bunkyo, Tokyo 113-0033, Japan}
\author{Shigehiro Nagataki}
\author{Katsuhiko Sato}
\affiliation{Department of Physics, School of Science, University of Tokyo, 7-3-1 Hongo, Bunkyo, Tokyo 113-0033, Japan}
\affiliation{Research Center for the Early Universe, School of Science, University of Tokyo, 7-3-1 Hongo, Bunkyo, Tokyo 113-0033, Japan}

\date{
\today
}

\begin{abstract}
We investigate the spectral shape of a high-energy diffuse photon emitted by evaporating primordial black holes (PBHs) in the Randall-Sundrum type II (RS2) braneworld.
In their braneworld scenario, the nature of small PBHs is drastically modified from the ordinary four-dimensional case for the following two reasons.
(i) dropping Hawking temperature, which equivalently lengthens the lifetime of the individual PBH due to the change of space-time topology and (ii) the effective increase of the total amount of PBHs caused by accretion during the earliest part of the radiation-dominated epoch, the brane high-energy phase.
From studies of the expected spectral shape and its dependence on braneworld parameters, we obtain two qualitatively distinctive possibilities of constraints on the braneworld PBHs from the observations of diffuse high-energy photon background.
If the efficiency of accretion in the high-energy phase exceeds a critical value, the existence of the extra dimension gives a more stringent upper bound on the abundance of PBHs than the 4D case and a small length scale for the extra dimension is favored.
On the contrary, in the case below the critical accretion efficiency, we find that the constraint on the PBH abundance can be relaxed by a few orders of magnitude in exchange for the existence of the large extra dimension; its size may be even bounded in the region above $ 10^{19} $ times 4D Planck length scale provided the rest mass energy density of the PBHs relative to energy density of radiation is actually larger than $ 10^{-27} $ (4D upper bound) at their formation time.
The above analytical studies are also confirmed numerically, and an allowed region for braneworld parameters and PBH abundance is clearly obtained.
\end{abstract}



\preprint{UTAP-449}
\preprint{RESCEU-20/03}

\maketitle

\section{INTRODUCTION}

It is widely believed that the local density contrast in the inflationary early Universe caused gravitational collapse, which resulted in the subsequent formation of plenty of mini black holes, the so-called primordial black holes (PBHs) \cite{Zel'dovichNovikov1966,Hawking1971,CarrHawking1974,Carr1975,Novikovetal1979}.
Such small black holes are thought to be only able to have formed in the early Universe.
There are even other mechanisms that could cause the PBH formation \cite{Bringmannetal2001,Blaisetal2003,Rubinetal20002001}.

Since Hawking discovered the possibility that black holes can emit particles while decreasing their masses \cite{Hawking1974,Page1976}, investigating the energy spectra of emission from PBHs has been one of the vital ways to obtain informations on the earliest era of the Universe.
Small black holes in the Universe radiate like blackbodies at every moment in their lives and leave traces in their relics.
In particular PBHs which have lifetimes around the present cosmological age have been emitting only massless particles, whose optical depths are much less than unity; they are in principle directly detectable.
Hence comparing predicted spectra with present observations, we can obtain information on the early Universe via the abundance of PBHs.
One of the earliest works that applied the theory of black hole evaporation into astrophysics was carried out by Page and Hawking \cite{PageHawking1976}, who considered the diffuse $ \gamma $-ray background to probe the number density of PBHs that have evaporated just now.
With the same motivation, a number of arguments on the amount of PBHs have been reported \cite{Carr1976,MacGibbonCarr1991,Barrauetal2003}.

More than two decades after Hawking's discovery, it was suggested by Randall and Sundrum \cite{RandallSundrum1999} that our Universe is a 4D Lorentz metric hypersurface, called the {\it brane}, embedded in $ Z_2 $ symmetric 5D anti--de Sitter (AdS) spacetime, called the {\it bulk}.
The so-called RS2 braneworld model, which we concentrate on in this paper, is one of the probable candidate models for our Universe they suggested.
It assumes that the number of branes is only one and the brane has positive tension.
In their scenario, ordinary matter is confined on the brane while 4D gravity is reproduced due to effective compactification.
The bulk has a typical length scale that originates from AdS curvature radius, $ l $.
This length scale draws a short distance boundary on the brane; deviation from ordinary gravity appears below it, while 4D Newton law is recovered as far as we consider a sufficiently larger scale than $ l $ \cite{RandallSundrum1999,SMS2000}.
The present table-top upper bound on the bulk AdS radius $ l $ was obtained by measuring short-distance gravitational force as $ l \lesssim 0.2~\mathrm{mm} $ \cite{Hoyleetal2001}.
In other words, there is a possibility that brane signature is directly seen by observing submillimeter scale gravitational interaction.

From such a point of view, a black hole, the most strongly gravitationally bounded object, is one of the most appropriate probes for submillimeter physics.
Actually, in the framework of the braneworld it has been revealed that the nature of black holes with small masses highly differs from the ordinary 4D one.
In particular small black holes with Schwarzschild radius $ r_\mathrm{S} \ll l $ behave as essentially 5D objects and their mass-radius relation is modified \cite{Emparanetal2000a,Wiseman2002,GiddingsThomas2002,CasadioMazzacurati2003,Kudohetal2003}.
On the other hand there have been discussions on the black hole formation in respect of the particle scattering processes \cite{GiddingsThomas2002,AhnCavaglia2003}, but we naturally expect that even such small black holes were produced by gravitational collapse in the early universe \cite{GCL2002a}.
Thus we have to revisit the particle emission process of primordial black holes taking into account how non-4D character modifies it.
In addition, the RS2 scenario implies a nonstandard high-energy regime before the standard radiation-dominated era \cite{Clineetal1999,Csakietal1999,Binetruyetal2000a,Binetruyetal2000b} (see, e.g., \cite{Langlois2003} and references therein).
It was recently argued by Guedens, Clancy, and Liddle and Majumdar \cite{GCL2002a,GCL2002b,Majumdar2003} that through the brane high-energy phase accretion of surrounding radiation onto PBHs is so efficient that they grow in their masses.
On the other hand, such accretion can be neglected as the Universe enters the standard radiation-dominated era.

Furthermore, taking into account the above five-dimensional signatures, Clancy, Guedens, and Liddle \cite{CGL2003} discovered that the shape of diffuse photon spectrum emitted from PBHs is drastically modified from that calculated by Page and Hawking in the 4D case, especially the peak energy can drop by three orders of magnitude.
Therefore it was revealed that we can obtain crucial information on the earliest universe in the braneworld scenarios by investigating those spectral shapes carefully.
However, what was obtained by Clancy {\it et al.} is only a rough shape of the photon spectrum.
Although they even estimated an upper bound on initial mass fraction of PBHs, it was only for one specific parameter set where the effect of accretion was ignored.
In other words they argued about a revised upper bound on the PBH abundance standing on the theory of 5D evaporating black holes, but 5D brane cosmology was not significantly taken into consideration.

On such a background our purpose in this paper is to reveal the detailed dependence of spectral shape of emitted photons on the size of extra dimension $ l $ and accretion efficiency $ F $.
For this purpose we first derive how the PBH mass spectrum is modified by accretion, which has not been carefully considered before.
Next we estimate the emitted photon spectrum taking the modified PBH mass spectrum into account, and compare them with the diffuse high-energy photon background spectrum observed by HEAO-1 A2 \cite{Gruber1992}, HEAO-1 A4 \cite{Kinzeretal1997}, SMM \cite{Kappadath1998}, COMPTEL \cite{Kappadath1998}, and EGRET \cite{Sreekumaretal1998}.
In the analysis, we particularly pay attention to the fact that the observed spectrum obeys nearly a power-law of index $ -1 $ in almost all energy regions.
It turns out later that this index determines a critical accretion efficiency.
Additionally we also concentrate on the origin of difference between the new 5D constraints and the old 4D one.
We find that there is a kind of critical size around $ l \approx 5 \times 10^{19} l_4 $ in the framework of evaporating PBHs.

For further analysis we even numerically study diffuse high-energy photon spectra for various possible braneworld parameter sets.
We use throughout this paper the newest observational cosmological parameters based on Wilkinson Microwave Anisotropy Probe (WMAP) observations \cite{Bennettetal2003}, which confirmed the preceding suggestions of a $ \Lambda $-dominated universe obtained from measurements on high-redshift type Ia supernovae \cite{Riessetal1998,Perlmutteretal1999}.
We also take into consideration precisely the effect of a mass decrease of PBHs in the evaporating process.
Hence expected shapes of diffuse photon spectrum from braneworld PBHs are obviously shown.

The contents of this paper are as follows.
First we briefly review in Sec.~\ref{sec:review} cosmology and behavior of evaporating black holes in the RS2 braneworld scenario.
In Sec.~\ref{sec:formulation}, formulations to describe PBHs in the brane universe are shown.
Those are basically according to the preceding papers \cite{GCL2002a,GCL2002b,Majumdar2003,CGL2003}.
After the theoretical preparations, we analyze general dependences of the PBH mass spectrum and resulting  photon spectrum on the braneworld parameters in Sec.~\ref{sec:analysis}.
In this section, our main purpose, new constraints on braneworld parameters and their implications are obtained.
Finally in Sec.~\ref{sec:numerical}, we show numerical results on the calculation of diffuse photon spectra for a wide range of braneworld parameters and refine and visualize our analysis in the previous section.
Conclusions and discussions are presented in Sec.~\ref{sec:conclusion}.

\section{\label{sec:review}
COSMOLOGY AND BLACK HOLES IN THE BRANEWORLD
}

\subsection{Brane cosmology}

Hereafter we basically use 4D natural units, that is $ \hbar = c = k_\mathrm{B} = 1 $ throughout this paper.
On the other hand 4D Planck scale quantities defined as $ M_4 $, $ l_4 $, $ t_4 $, and $ T_4 $ usually appear in expressions because the braneworld scenario, which we consider here, is a higher-dimensional unified theory.

We introduce a five-dimensional fundamental scale $ \kappa_5 $ of mass dimension $ -3/2 $, which appears in the 5D Einstein equation as
\begin{equation}
G_{MN}^{(5)} = -\Lambda_5 g_{MN}^{(5)} + \kappa_5^2 T_{MN}^{(5)}.
\end{equation}
Here $ g_{MN}^{(5)} $ is 5D metric, $ G_{MN}^{(5)} $ and $ T_{MN}^{(5)} $ are the 5D Einstein tensor and energy-momentum tensor, respectively.
$ \Lambda_5 $ is the negative cosmological constant in the bulk.
Expansion-law on the brane is determined by this five-dimensional equation with an appropriate boundary condition on the brane, so-called Israel's junction condition.
In this framework, brane cosmological expansion is modified from the 4D case due to the existence of typical bulk length scale, the AdS curvature radius $ l $, and energy scale on the brane, the brane tension $ \lambda $ \cite{SMS2000,Clineetal1999,Csakietal1999,Binetruyetal2000a,Binetruyetal2000b}.
The AdS curvature radius is defined by bulk energy scales as
\begin{equation}
l = \sqrt{- \frac{6}{\kappa_5^2 \Lambda_5}}.
\end{equation}
The quantity $ l $ expresses how much the bulk has physical volume, hence according to the size of $ l $ we mention whether the extra dimension is large or small.

Using the 5D Einstein equation in the bulk and the boundary condition on the brane under $ Z_2 $ symmetry, one can obtain the 4D Friedmann equation on the brane in a modified form:
\begin{equation}
H^2 = \frac{8 \pi}{3 M_4^2} \left[\rho \left(1 + \frac{\rho}{2 \lambda}\right) + \rho_\mathrm{KK}\right] + \frac{\Lambda_4}{3} - \frac{K}{a^2} \label{eq:Friedmann1},
\end{equation}
where $ a $ is the scale factor, $ H $ is the Hubble constant, $ \rho $ and $ \rho_\mathrm{KK} $ are the energy density of ordinary matter field and of the dark radiation, respectively, and $ K $ is the spatial curvature of the brane.
$ M_4 $ and $ \Lambda_4 $ are the induced Planck mass and cosmological constant on the brane related to 5D quantities as
\begin{equation}
\Lambda_4 = \frac{\kappa_5^2}{2} \left(\Lambda_5 + \frac{\kappa_5^2 \lambda^2}{6}\right), ~~~ M_4^2 = \frac{48 \pi}{\kappa_5^4 \lambda}.
\end{equation}
It is needed for these quantities to match observed 4D values.
The energy conservation law is the same form as the 4D one:
\begin{equation}
\dot{\rho} + 3 H (\rho + p) = 0
\end{equation}
(see, e.g., \cite{Langlois2003} and references therein for a review).

Here we evaluate the terms in the above modified Friedmann equation.
As we are interested in a case which is connected with the well-established standard 4D universe, each term is bounded or even determined by current observations.
As for the dark radiation term, it was argued that this quantity must be sufficiently smaller than ordinary radiation at the time of nucleosynthesis \cite{Binetruyetal2000b,BarrowMaartens2002}, thus we drop this term.
Next the observations of WMAP imply that spatial curvature of the brane is small \cite{Bennettetal2003}, hence we also set $ K = 0 $.
Though WMAP also revealed that the 4D cosmological constant $ \Lambda_4 $ takes a significant role in the present Universe, it can be ignored in the early Universe.
We require $ \Lambda_4 $ to vanish except when the present epoch of the Universe is considered.
Therefore one finds formal Friedmann equation (\ref{eq:Friedmann1}) results in the following form:
\begin{equation}
H^2 = \frac{8 \pi}{3 M_4^2} \rho \left(1+\frac{\rho}{2 \lambda}\right) + \frac{\Lambda_4}{3}. \label{eq:Friedmann2}
\end{equation}
The above expression differs from the ordinary 4D one with respect to the existence of a $ \rho $-square term.
This term is only relevant when energy density is sufficiently higher than the brane tension, namely in earliest era of the Universe.
Then we can split the cosmological evolution in the early Universe into two phases; the earlier one is the nonstandard high energy regime in which evolution of the scale factor $ a(t) $ obeys $ a(t) \propto t^{1/4} $, and the latter is the standard radiation-dominated universe, where $ a(t) \propto t^{1/2} $.
The transition time between them is $ t_\mathrm{c} \equiv l/2 $.

Physical quantities governing cosmological expansion should be determined by observation.
Thus the procedure of determining the profile of scale factor $ a(t) $ is as follows.
To begin with, we normalize the scale factor so that its present value becomes unity: $ a(t_0)=1 $.
Then we integrate the Friedmann equation to the past using WMAP data, such as $ t_0=13.7~\text{Gyr} $, $ h=0.71 $, $ \Omega_{\mathrm{m},0}h^2=0.135 $, $ \Omega_{\Lambda,0}=0.73 $, and $ z_\mathrm{eq}=3233 $.
On the other hand, as for radiation energy density, observed data taken by Cosmic Background Explorer (COBE)  \cite{Matheretal1994} is used as the present value: $ \Omega_{\mathrm{r},0} h^2 = 4.15 \times 10^{-5} $.
Then all we have to do is go back to the time when $ \Omega_\mathrm{m} \gg \Omega_\Lambda $ holds and connects to the Einstein-de Sitter universe.
The choice of ``connection time,'' $ t_\Lambda $, is artificial, and hence eventually may cause an error in other quantities; we carefully chose an appropriate $ t_\Lambda $ so that the time of matter-radiation equality $ t_\mathrm{eq} $, which is determined by $ a_\mathrm{eq} $ with $ t_\Lambda $, has no meaningful numerical error.
As a consequence the profile of scale factor is determined as
\begin{subequations}
\begin{eqnarray}
a(t)
 &=& a_\mathrm{eq} \frac{t^{1/4} t_\mathrm{c}^{1/4}}{t_\mathrm{eq}^{1/2}} ~~~ \text{for}~t \ll t_\mathrm{c} \\
 &=& a_\mathrm{eq} \frac{t^{1/2}}{t_\mathrm{eq}^{1/2}} ~~~~~~~~ \text{for}~t_\mathrm{c} \ll t \leq t_\mathrm{eq} \\
 &=& a_\mathrm{eq} \frac{t'^{2/3}}{t_\mathrm{eq}'^{2/3}} ~~~~~~~~ \text{for}~t_\mathrm{eq} \leq t \leq t_\Lambda,
\end{eqnarray}
\end{subequations}
where the value of $ t_\mathrm{eq} $ is calculated as 72.6~kyr and a prime denotes redefinition of time origin ($ t' = t'_\mathrm{eq} $ when $ t = t_\mathrm{eq} $).
Note that although we often make use of these asymptotic forms there is even an analytic solution through the whole radiation-dominated era:
\begin{equation}
a(t) = a_\mathrm{eq} \frac{t^{1/4} (t+t_\mathrm{c})^{1/4}}{t_\mathrm{eq}^{1/2}}.
\end{equation}
It is also utilized to describe near the transition time $ t_\mathrm{c} $.

Using the above profile of $ a(t) $, other cosmological quantities, energy density $ \rho(t) $, Hubble radius $ R_\mathrm{H}(t) $, and horizon mass $ M_\mathrm{H}(t) $ are calculated.
We are particularly interested in their expressions in the early Universe:
\begin{subequations}
\begin{eqnarray}
\rho(t) &=& \frac{3 M_4^2}{32 \pi t_\mathrm{c} t}, ~~~ R_\mathrm{H}(t) ~=~ 4 t, \nonumber \\
M_\mathrm{H}(t) &=& 8 M_4^2 \frac{t^2}{t_\mathrm{c}} ~~~~~ \text{for}~t \ll t_\mathrm{c}, \label{eq:highenergyphase} \\
\rho(t) &=& \frac{3 M_4^2}{32 \pi t^2}, ~~~~ R_\mathrm{H}(t) ~=~ 2 t, \nonumber \\
M_\mathrm{H}(t) &=& M_4^2 t ~~~~~~~~~ \text{for}~t \gg t_\mathrm{c}.
\end{eqnarray}
\end{subequations}

\subsection{Brane black holes}

In the braneworld scenario, a massive object localized on the brane with typical radius $ r \ll l $ should see the background space-time as effectively flat 5D Minkowski space \cite{Wiseman2002,GiddingsThomas2002,CasadioMazzacurati2003,Kudohetal2003}.
Therefore a sufficiently small black hole is no longer an ordinary 4D one, but can be approximately described as 5D Schwarzschild solution.
In terms of general spherical coordinate, the 5D Schwarzschild metric is \cite{MyersPerry1986}
\begin{equation}
\mathrm{d}s_5^2 = -C(r) \mathrm{d}t^2 + \frac{1}{C(r)} \mathrm{d}r^2 + r^2 \mathrm{d}\Omega_3^2
\end{equation}
with
\begin{equation}
C(r) = 1 - \left(\frac{r_\mathrm{S}}{r}\right)^2,
\end{equation}
where $ r_\mathrm{S} $ is the Schwarzschild radius and $ \mathrm{d}\Omega_D $ is the line element on unit $ D $-sphere.
The Schwarzschild radius in the braneworld is given as \cite{Emparanetal2000a}
\begin{equation}
r_\mathrm{S} = \sqrt{\frac{8}{3 \pi}} \left(\frac{l}{l_4}\right)^{1/2} \left(\frac{M}{M_4}\right)^{1/2} l_4. \label{eq:radius}
\end{equation}
The induced 4D metric near the event horizon is obtained by projecting the bulk 5D metric onto brane,
\begin{equation}
\mathrm{d}s_4^2 = -C(r) \mathrm{d}t^2 + \frac{1}{C(r)} \mathrm{d}r^2 + r^2 \mathrm{d}\Omega_2^2 \label{eq:metric}.
\end{equation}
On the other hand general $ D $-dimensional Hawking temperature of a Schwarzschild black hole is given in terms of its radius as \cite{MyersPerry1986,GiddingsThomas2002}
\begin{equation}
T_\mathrm{BH} = \frac{D-3}{4 \pi r_\mathrm{S}} \label{eq:Hawkingtemperature_D}.
\end{equation}
Combining these expressions, one finds the modified mass to Hawking temperature relation,
\begin{equation}
T_\mathrm{BH} = \sqrt{\frac{3}{32 \pi}} \left(\frac{l}{l_4}\right)^{-1/2} \left(\frac{M}{M_4}\right)^{-1/2} T_4. \label{eq:Hawkingtemperature}
\end{equation}
While ordinary 4D expressions are
\begin{eqnarray}
C(r)(\text{4D}) &=& 1-\frac{r_\mathrm{S}}{r}, \\
r_\mathrm{S}(\text{4D}) &=& 2 \frac{M}{M_4} l_4, \\
T_\mathrm{BH}(\text{4D}) &=& \frac{1}{8 \pi} \left(\frac{M}{M_4}\right)^{-1} T_4.
\end{eqnarray}

As far as we consider such small black holes for which the above 5D approximations are valid, the temperature of a black hole with some fixed mass $ M $ is to be in general lower than in the 4D universe.
We see in the next section how this change affects the spectra of particles emitted from black holes.

\section{\label{sec:formulation}
FORMULATIONS
}

\subsection{PBH formation}

It has been argued that in the early Universe PBHs are formed due to gravitational collapse caused by density perturbation.
Though there are theoretical difficulties on the process of gravitational collapse,\footnote{
Actually there is a paper showing critical phenomena in the formation of PBHs \cite{NiemeyerJedamzik1999}.
However, it is still reasonable to take $ f = \mathcal{O}(1) $ in Eq.~(\ref{eq:initialmass}).
} we here simplify the situation; when a superhorizon scale density perturbation with adequate amplitude enters the Hubble horizon at $ t = t_\mathrm{i} $, it instantly forms a black hole with some fraction of horizon mass.
Hence the formation mass $ M_\mathrm{i} $ is given by
\begin{equation}
M_\mathrm{i} = f M_\mathrm{H}(t_\mathrm{i}) \label{eq:initialmass},
\end{equation}
where $ f $ expresses the fraction of horizon mass collapsing into the black hole.
It can be confirmed that $ f $ is order unity by virtue of Jeans analysis and that such a black hole has a Schwarzschild radius nearly covering the Hubble horizon \cite{GCL2002a}.

\subsection{Accretion on PBHs}

Radiation field localized on the brane sees a black hole approximately as a disk-shaped object with radius $ r_{\mathrm{S},\mathrm{eff}} $, where $ r_{\mathrm{S},\mathrm{eff}} $ is the effective $ D $-dimensional Schwarzschild radius given by \cite{Emparanetal2000b}
\begin{equation}
r_{\mathrm{S},\mathrm{eff}} = \left(\frac{D-1}{2}\right)^{1/(D-3)} \left(\frac{D-1}{D-3}\right)^{1/2} r_\mathrm{S}.
\end{equation}
Consider an infinitesimal time interval $ [t,t+\mathrm{d}t] $.
Then the corresponding gain of black hole mass due to absorption of surrounding radiation is described as \cite{GCL2002b,Majumdar2003}
\begin{equation}
\frac{\mathrm{d}M}{\mathrm{d}t} = F \pi r_{\mathrm{S},\mathrm{eff}}^2 \rho(t),
\end{equation}
where $ \rho(t) $ is energy density of radiation.
Here we introduced a parameter $ F $, which determines accretion efficiency.
If the geometrical optics approximation holds well, namely an incidental relativistic particle can be regarded as a collisionless point particle and its spin can be ignored, $ F $ should be almost unity, but otherwise smaller.

Substituting relevant expressions into the above differential equation and integrating it, we obtain the mass evolution relation in the brane high energy phase as \cite{GCL2002b,Majumdar2003}
\begin{equation}
M(t) \approx \left(\frac{t}{t_\mathrm{i}}\right)^{2 F/\pi} M_\mathrm{i} ~~~ \mbox{for}~t \lesssim t_\mathrm{c}.
\end{equation}
We define by the above formula $ M_\mathrm{c} \equiv M(t_\mathrm{c}) $.

The effect of evaporation was ignored in the above derivation.
Thus one may suspect the above expression is not valid when accretion does not work, say $ F=0 $.
However, evaporation in such a short period naturally has little significance for long-life PBHs which we will consider later; we can safely make use of the above expression regardless of the value of $ F $.

\subsection{Evaporation}

The differential particle emission rate from a $ D $-dimensional static noncharged black hole with mass $ M $ is given by Hawking's formula:
\begin{equation}
\mathrm{d}\frac{\mathrm{d}N_j}{\mathrm{d}t} = g_j \frac{\sigma_j(E)}{\exp(E/T_\mathrm{BH}) \pm 1} \frac{\mathrm{d}k^{D-1}}{(2 \pi)^{D-1}},
\end{equation}
where the subscript $ j $ indicates particle species, $ g_j $ is the number of internal degrees of freedom of the particle, $ E = \sqrt{k^2 + m_j^2} $ is the relativistic energy including rest mass $ m_j $, $ \sigma_j $ is the absorption cross section, the sign in the denominator expresses which statistics the particle obeys, and $ T_\mathrm{BH} $ is the Hawking temperature related to the black hole Schwarzschild radius given in Eq.~(\ref{eq:Hawkingtemperature_D}).
Using the above formula under relativistic limit, the total mass decreasing rate of a spherically symmetric black hole results in
\begin{eqnarray}
\frac{\mathrm{d}M}{\mathrm{d}t}
 &=& -\sum_j \int_0^\infty E \frac{\mathrm{d}^2N_j}{\mathrm{d}E \mathrm{d}t} \mathrm{d}E \nonumber \\
 &=& -\sum_j \int_0^\infty g_j \frac{\sigma_j(E)}{\exp(E/T_\mathrm{BH}) \pm 1} \frac{\Omega_{D-2} E^{D-1} \mathrm{d}E}{(2\pi)^{D-1}} \nonumber \\
 &\approx& -\sum_j \int_0^\infty g_j \frac{A_{\mathrm{eff},D}/4}{\exp(E/T_\mathrm{BH}) \pm 1} \frac{\Omega_{D-2} E^{D-1} \mathrm{d}E}{(2\pi)^{D-1}} \nonumber \\
 &\approx& -g_D \sigma_D A_{\mathrm{eff},D} T_\mathrm{BH}^D.
\end{eqnarray}
Here we introduced some quantities following \cite{GCL2002a};
$ g_D $ is an effective degree of freedom given by
\begin{equation}
g_D = g_{D,\mathrm{boson}} + \frac{2^{D-1}-1}{2^{D-1}} g_{D,\mathrm{fermion}},
\end{equation}
$ \sigma_D $ is the $ D $-dimensional Stefan-Boltzmann constant
\begin{equation}
\sigma_D = \frac{\Omega_{D-2}}{4 (2 \pi)^{D-1}} \Gamma(D) \zeta(D),
\end{equation}
and $ A_{\mathrm{eff},D} $ is the $ (D-2) $-dimensional effective surface area
\begin{equation}
A_{\mathrm{eff},D} = \Omega_{D-2} r_{\mathrm{S},\mathrm{eff},D}^{D-2},
\end{equation}
where $ \Omega_{D-2} $ is the area of the $ (D-2) $-dimensional unit sphere, and $ \Gamma(D) $ and $ \zeta(D) $ are gamma and zeta functions, respectively.

To investigate the evaporation process of black holes in the braneworld, one should consider degrees of freedom on the brane and in the bulk separately because the former takes 3D phase space and 2D black hole surface area while the latter takes 4D phase space and 3D surface area.
Thus the mass shedding formula takes the form below:
\begin{equation}
\frac{\mathrm{d}M}{\mathrm{d}t}
 = -g_\mathrm{brane} \tilde{A}_{\mathrm{eff},4} T_\mathrm{BH}^4
 - g_\mathrm{bulk} A_{\mathrm{eff},5} T_\mathrm{BH}^5.
\end{equation}
Note that we must take $ \tilde{A}_{\mathrm{eff},4} = \pi r_{\mathrm{S},\mathrm{eff},5}^2 $ since the Schwarzschild radius is determined by the five-dimensional metric.

From the above arguments, we obtained the fundamental differential equation which describes mass-loss rate of black holes.
Thus by integrating it, we know how the mass of a black hole decreases and therefore how the Hawking temperature rises during its lifetime.
(A possible deviation from the above formula is discussed in \cite{CasadioHarms2002}.)

One theoretical uncertainty is on the treatment of effective degrees of freedom for high temperature black holes with $ T_\mathrm{BH} \gtrsim m_\pi $, where $ m_\pi \simeq 135~\mathrm{MeV} $ is the mass of pion as the lightest hadron.
However, the heat-up rate of a black hole in its early phase is so slow and once $ T_\mathrm{BH} $ gets as high as $ m_\pi $, a black hole ``blows up'' and ends its life almost instantly.
Thus in the following analysis, we can assume an evaporating black hole holds approximately the same temperature as the initial value through its lifetime, and hence the effective total degrees of freedom, $ g_\mathrm{eff} $, is constant.
With low temperature assumption ($ T_\mathrm{BH} \lesssim 100~\text{MeV} $), we can conclude that the relevant particle species to calculate $ g_\mathrm{eff} $ for those PBHs surviving until present epoch are massless species and at most lightest (anti-)leptons, i.e., photon, neutrinos, and electron/positron.
In the last part of this section, we will check the consistency of the low temperature assumption and, for concreteness, estimate the fraction of photon energy emitted after the assumption breaks down.
On the other hand we still do not have enough knowledge to precisely determine the spin-dependent characters of particle emission.
We here take the values obtained in the 4D case by Page \cite{Page1976}.
As a consequence, the mass shedding rate is given as
\begin{equation}
\frac{\mathrm{d}M}{\mathrm{d}t} = -\frac{g_\mathrm{eff}}{2} \left(\frac{l}{l_4}\right)^{-1} \left(\frac{M(t)}{M_4}\right)^{-1} \frac{M_4}{t_4}. \label{eq:massloss}
\end{equation}
We here cite the value of the effective degree of freedom $ g_\mathrm{eff} $ in \cite{GCL2002a}:
\begin{eqnarray}
g_\mathrm{eff}
 &\approx& 0.023 ~~~ \text{for only massless} \nonumber \\
 &\approx& 0.032 ~~~ \text{for massless and electron/positron}. \nonumber
\end{eqnarray}
Difference between them is almost negligible.
The solution for Eq.~(\ref{eq:massloss}) is easily obtained as
\begin{equation}
M(t) = \left[\left(\frac{M_\mathrm{c}}{M_4}\right)^2 - g_\mathrm{eff} \left(\frac{l}{l_4}\right)^{-1} \left(\frac{t-t_\mathrm{c}}{t_4}\right) \right]^{1/2} M_4.
\end{equation}
Here we took into account the appropriate initial condition that $ M=M_\mathrm{c}~\text{at}~t=t_\mathrm{c} $.
From this formula the lifetime of black hole $ t_\mathrm{evap} $ with initial mass $ M_\mathrm{c} $ can be estimated as
\begin{equation}
t_\mathrm{evap} = g_\mathrm{eff}^{-1} \left(\frac{l}{l_4}\right) \left(\frac{M_\mathrm{c}}{M_4}\right)^2 t_4 + t_\mathrm{c}. \label{eq:lifetime}
\end{equation}
Note that the lifetime for a PBH hotter than 100~MeV is significantly shortened and the above expression cannot be applied any longer.

We derive some typical values of black hole mass and temperature.
First of all, PBHs just formed at the end of the brane high-energy phase have mass $ M_\mathrm{c} \sim l $ and radius $ r_\mathrm{S} \sim l $.
Then $ l \sim 10^{20} l_4 $ is needed to lower the temperature of such PBHs around 100~MeV and hence above $ 10^{20} l_4 $ the preceding estimation of lifetime is valid; it gives $ t_\mathrm{evap} \sim  (l/10^{20}l_4)^3 t_0 $.
Thus we find that PBHs evaporating at the present epoch enjoy five-dimensionality only with $ l $ larger than $ 10^{20}l_4 $.
Later on it will also be revealed that PBHs relevant for our analysis are only those with a lifetime $ \sim t_0 $.
If $ l $ falls below $ 10^{20} l_4 $ we cannot see the difference from the 4D case.
Therefore we only have to pay attention to the case $ l \gtrsim 10^{20} l_4 $ throughout this paper.

For those black holes with lifetime $ t_\mathrm{evap}=t_0=13.7~\mathrm{Gyr} $, their initial mass $ M_\mathrm{c}^* $ is calculated as
\begin{equation}
M_\mathrm{c}^* = 3.0 \times 10^{9} \left(\frac{l}{10^{31} l_4}\right)^{-1/2}~\mathrm{g}. \label{eq:M_c^*}
\end{equation}
Thus corresponding initial temperature is
\begin{equation}
T_\mathrm{BH}^* = 57 \left(\frac{l}{10^{31} l_4}\right)^{-1/4}~\mathrm{keV}. \label{eq:Hawkingtemperature^*}
\end{equation}
With $ l $ in the typical range $ 10^{20} l_4 \text{--} 10^{31} l_4 $ of our interest, $ T_\mathrm{BH}^* $ is
\begin{equation}
57~\mathrm{keV} < T_\mathrm{BH}^* < 32~\mathrm{MeV}, \label{eq:Hawkingtemperature^*region}
\end{equation}
hence the consistency of low temperature approximation for those PBHs is confirmed.
Furthermore, since the black hole mass corresponding to $ T_\mathrm{BH} \sim 100~\text{MeV} $ is $ M|_{T_\mathrm{BH}=100~\text{MeV}} \sim 10^4 (l/10^{31}l_4)^{-1}~\text{g} $, the ratio $ M|_{T_\mathrm{BH}=100~\text{MeV}}/M_\mathrm{c}^* $ is much less than unity except $ l $ is very close to $ 10^{20} l_4 $.
Therefore we also confirm that the fraction of photon energy emitted after the Hawking temperature rises above the cutoff temperature $ m_\pi $ is negligible for those long-life PBHs.

Summarizing the previous sections, our formula to describe the time evolution of black hole mass is obtained as
\begin{subequations}
\begin{eqnarray}
M(t)
 &=& \left(\frac{t}{t_\mathrm{i}}\right)^{2 F/\pi} M_\mathrm{i} ~~~ \mbox{for} ~ t \leq t_\mathrm{c} \label{eq:massgrowth} \\
 &=& \left[\left(\frac{M_\mathrm{c}}{M_4}\right)^2 - g_\mathrm{eff} \left(\frac{l}{l_4}\right)^{-1} \left(\frac{t-t_\mathrm{c}}{t_4}\right) \right]^{1/2} M_4 \nonumber \\
 & & ~~~~~~~~~~~~~~~~~~~~ \text{for}~t \geq t_\mathrm{c}, \label{eq:massdepletion}
\end{eqnarray}
\end{subequations}
where $ M_\mathrm{c} = M(t_\mathrm{c}) $.

\subsection{PBH abundance}

In order to evaluate the fraction of energy density in the form of PBHs to that of radiation, we take an usual notation
\begin{equation}
\alpha_t(M_\mathrm{i}) \equiv \frac{\rho_{\mathrm{PBH},M_\mathrm{i}}(t)}{\rho_\mathrm{rad}(t)},
\end{equation}
where $ \rho_{\mathrm{PBH},M_\mathrm{i}} $ is the mass density of PBHs with formation mass on the order of $ M_\mathrm{i} $, and $ \rho_\mathrm{rad} $ is the radiation energy density.
In this paper we only use $ \alpha_\mathrm{i} \equiv \alpha_{t_\mathrm{i}} $, where $ t_\mathrm{i} $ is given for $ M_\mathrm{i} $ by solving Eq.~(\ref{eq:initialmass}).
$ \alpha_t $ is related to $ \alpha_\mathrm{i} $ via cosmological scale factor $ a(t) $ as
\begin{equation}
\alpha_t(M_\mathrm{i}) = \alpha_\mathrm{i}(M_\mathrm{i}) \frac{a(t)}{a(t_\mathrm{i})}.
\end{equation}
$ \alpha_\mathrm{i} $ is an important quantity for the reason that it is directly connected with the primordial density perturbation.
The upper limit on $ \alpha_\mathrm{i} $, which we are seeking, is written as $ \mathrm{LIM_i} $.

Using the above definitions, the number density of PBHs with formation mass on the order of $ M_\mathrm{i} $ is given as
\begin{equation}
n(M_\mathrm{i}) = \alpha_\mathrm{i}(M_\mathrm{i}) \frac{a(t)}{a(t_\mathrm{i})} \frac{\rho_\mathrm{rad}(t)}{M_\mathrm{i}} \theta(t_\mathrm{evap} - t), \label{eq:initialspectrum_def}
\end{equation}
where $ t_\mathrm{evap} $ is corresponding lifetime and $ \theta(t) $ is the theta function.

We note that the above number density $ n(M_\mathrm{i}) $ is a finite quantity although it apparently depends on specific mass $ M_\mathrm{i} $; this is because the mass fraction $ \alpha_\mathrm{i}(M_\mathrm{i}) $ is a quantity for PBHs with mass {\it on the order of} $ M_\mathrm{i} $.
The relation between $ n(M_\mathrm{i}) $ and the differential mass spectrum $ \mathrm{d}n_\mathrm{PBH}/\mathrm{d}M_\mathrm{i} $ can be written as
\begin{equation}
n(M_\mathrm{i}) = \int_{0.1 \beta M_\mathrm{i}}^{\beta M_\mathrm{i}} \mathrm{d}M_\mathrm{i} \frac{\mathrm{d}n_\mathrm{PBH}}{\mathrm{d}M_\mathrm{i}} \sim M_\mathrm{i} \frac{\mathrm{d}n_\mathrm{PBH}}{\mathrm{d}M_\mathrm{i}}, \label{eq:numberdensity}
\end{equation}
where $ \beta $ is a constant on the order of unity.
Therefore some constraint on $ n(M_\mathrm{i}) $ leads to an equivalent constraint on the mass spectrum $ \mathrm{d}n_\mathrm{PBH}/\mathrm{d}M_\mathrm{i} $.

Throughout this paper, we only consider the number density $ n(M_\mathrm{i}) $ rather than the fundamental mass spectrum.
This treatment, although it has been usually employed in arguments on PBHs, may cause differences in the results due to the uncertainty in the definition of $ n(M_\mathrm{i}) $, however, our conclusions will not be qualitatively changed.
In the following discussion we often call number density $ n(M_\mathrm{i}) $ ``mass spectrum'' because those two quantities are essentially equivalent.

\subsection{Diffuse photon}

As a consequence of Hawking radiation, PBHs emit particles at every moment and then reduce their masses obeying the formula (\ref{eq:massdepletion}).
Particularly we are interested in photons emitted between the decoupling time $ t_\mathrm{dec} $ and the present epoch $ t_0 $ because they have never strongly interacted with matter and hence its original spectral shape should be preserved in the observable high-energy diffuse photon background.
But unfortunately it is understood that we cannot see its signature in the present observations; all we can do is to derive expected spectrum and set the upper bound on PBH abundance.

As for photons, the relevant differential emission rate of Hawking radiation is
\begin{equation}
\frac{\mathrm{d}^2N}{\mathrm{d}E \mathrm{d}t} = \frac{\sigma(E)}{\exp(E/T_\mathrm{BH})-1} \frac{E^2}{\pi^2}.
\end{equation}
Instead of integrating it, we use an approximated expression for the emission rate of energetic photons:
\begin{equation}
\frac{\mathrm{d}N}{\mathrm{d}t} \approx E \frac{\mathrm{d}^2N}{\mathrm{d}E \mathrm{d}t}.
\end{equation}
Thus total emission rate of energy density via photons with energy $ E $ is given as
\begin{equation}
\frac{\mathrm{d}U(E)}{\mathrm{d}t} \approx \int \mathrm{d}n(M_\mathrm{i}) E^2 \frac{\mathrm{d}^2N}{\mathrm{d}E \mathrm{d}t}.
\end{equation}
In the cosmological context, photon energy $ E $ emitted at time $ t $ suffers from redshift.
Thus the observed energy of photons emitted at time $ t $ with energy $ E $ is
\begin{equation}
E_0 = \frac{a(t)}{a(t_0)} E
\end{equation}
at present.
Additionally, the energy density of photons is also reduced by a factor of $ a^{-4} $.
Therefore resulting energy density $ U_0(E_0) $ observable now is expressed as follows:
\begin{eqnarray}
U_0(E_0)
 &=& \int_{t_\mathrm{dec}}^{t_0} \mathrm{d}t \frac{\mathrm{d}U(E)}{\mathrm{d}t} \frac{a^4(t)}{a^4(t_0)} \nonumber \\
 &=& \int \mathrm{d}M_\mathrm{i} \frac{\mathrm{d}n(M_\mathrm{i})}{\mathrm{d}M_\mathrm{i}} \int_{t_\mathrm{dec}}^{t_\mathrm{evap}} \mathrm{d}t E^2 \frac{\mathrm{d}^2N}{\mathrm{d}E \mathrm{d}t} \frac{a^4(t)}{a^4(t_0)}. \nonumber \\
 & & \label{eq:U_0}
\end{eqnarray}
Here the effect of absorption during propagation in the extra-galactic field is ignored because it should not cause significant deviation in the energy range of our interest.

Since the energy spectrum of emitted particles at a moment shows a sharp peak, it is generally expected that the spectrum of $ U_0 $ has two peaks, or at least edges.
One, the high-energy peak, is around the temperature of PBHs with lifetime $ t_0 $.
The other, the low-energy peak, is also determined by the temperature of PBHs with lifetime $ t_\mathrm{dec} $ but its location is redshifted by a factor of $ a(t_\mathrm{dec})/a(t_0) $.
The high-energy peak takes a crucial role in the following discussions, but the low-energy peak is less important.

When we are to compare the calculation with the observed spectrum, an adequate quantity is the surface brightness,
\begin{equation}
I(E_0) = \frac{c}{4\pi} \frac{U_0(E_0)}{E_0}, \label{eq:I}
\end{equation}
which has units of $ \mathrm{keVcm^{-2}sr^{-1}s^{-1}keV^{-1}} $.
From now on, ``(diffuse) photon spectrum'' is used for this quantity throughout this paper.

\section{\label{sec:analysis}
ANALYSIS ON SPECTRA
}

In this section, we investigate two different kinds of spectra, the PBH mass spectrum and the diffuse photon spectrum.

To describe the time evolution of mass spectrum, we have to consider two kinds of masses, $ M_\mathrm{i} $ and $ M_\mathrm{c} $.
Although we defined the former as the mass at the initial time of individual PBHs, it appears natural to treat the latter as an {\it initial} value of successive time evolution.
Therefore in the following sections we basically call $ M_\mathrm{c} $ and $ M_\mathrm{i} $ ``initial mass'' and ``formation mass,'' respectively.

\subsection{Relation between PBH mass spectrum and diffuse photon spectrum}

Since we are mainly interested in PBHs having evaporated away until now, we consider light PBHs formed in the brane high energy phase, where the approximation of $ t_\mathrm{i} \ll t_\mathrm{c} $ holds well.
From Eqs.~(\ref{eq:highenergyphase}), (\ref{eq:initialmass}), and (\ref{eq:initialspectrum_def}), the number density of PBHs with formation mass $ M_\mathrm{i} $ is obtained as
\begin{eqnarray}
n(M_\mathrm{i})|_{t=t_\mathrm{i}}
 &=& \alpha_\mathrm{i}(M_\mathrm{i}) \frac{\rho_\mathrm{rad}(t_\mathrm{i})}{M_\mathrm{i}} \nonumber \\
 &=& \frac{3 M_4^3}{4 \pi} \alpha_\mathrm{i}(M_\mathrm{i}) f^{1/2} l^{-3/2} M_\mathrm{i}^{-3/2}. \label{eq:initialspectrum}
\end{eqnarray}
We should recall that this quantity $ n|_{t=t_\mathrm{i}} $ is not a spectrum at a specific time but only a collection of the values of number densities evaluated at the formation time of each PBH.
Thus the above mass spectrum is distorted by subsequent accretion; the mass growth ratio for individual PBHs differs according to their formation mass $ M_\mathrm{i} $.
Therefore resulting mass $ M_\mathrm{c} $ after accretion is obtained from Eqs.~(\ref{eq:highenergyphase}), (\ref{eq:initialmass}), and (\ref{eq:massgrowth}) as
\begin{eqnarray}
M_\mathrm{c}
 &=& \left(\frac{t_\mathrm{c}}{t_\mathrm{i}}\right)^{2 F/\pi} M_\mathrm{i} \nonumber \\
 &=& \left(4 M_4^2 \frac{f l}{M_\mathrm{i}}\right)^{F/\pi} M_\mathrm{i}.
\end{eqnarray}
Thus in turn the following relation holds:
\begin{equation}
M_\mathrm{i} = \left(4 M_4^2 f l\right)^{-F/(\pi-F)} M_\mathrm{c}^{\pi/(\pi-F)}. \label{eq:M_i}
\end{equation}
Combining Eqs.~(\ref{eq:initialspectrum}) and (\ref{eq:M_i}), we obtain the {\it initial} PBH mass spectrum at the end of accretion as\footnote{
Because $ n $ is an integrated quantity, Jacobian only contributes as a factor of order unity in the final form and we omit it for simplicity.
From Eq.~(\ref{eq:numberdensity}), one can see that this factor takes the form of $ \pi/(\pi-F) $.
}
\begin{eqnarray}
n(M_\mathrm{c})|_{t=t_\mathrm{c}}
 &\simeq& \frac{a(t_\mathrm{i})^3}{a(t_\mathrm{c})^3} n(M_\mathrm{i})|_{t=t_\mathrm{i}} \nonumber \\
 &=& \frac{3 M_4^{9/4}}{2^{7/2} \pi} \alpha_\mathrm{i}(M_\mathrm{i}) f^{1/8} l^{-15/8} M_\mathrm{i}^{-9/8} \nonumber \\
 &=& \frac{3 M_4^{9/4}}{2^{7/2} \pi} \alpha_\mathrm{i}[M_\mathrm{i}(M_\mathrm{c})] \nonumber \\
 & & \times (4 M_4^2)^{\F} f^{1/8+\F} l^{-15/8+\F} M_\mathrm{c}^{-9/8-\F}, \nonumber \\
 & & 
\end{eqnarray}
where a new accretion parameter $ \F \equiv 9F/(8\pi-8F) $ was introduced for convenience.
Though there remains $ M_\mathrm{i} $ as the argument of $ \alpha_\mathrm{i} $, it is explicitly determined by $ M_\mathrm{c} $ using Eq.~(\ref{eq:M_i}).
Finally, mass spectrum at any later time is given by the following form:
\begin{eqnarray}
n(M_\mathrm{c})
 &=& \frac{a(t_\mathrm{c})^3}{a(t)^3} n(M_\mathrm{c})|_{t=t_\mathrm{c}} \nonumber \\
 &=& \frac{3 M_4^{9/4}}{2^{17/4} \pi} \frac{a_\mathrm{eq}^3}{a(t)^3 t_\mathrm{eq}^{3/2}} \alpha_\mathrm{i}[M_\mathrm{i}(M_\mathrm{c})] \nonumber \\
 & & \times (4 M_4^2)^{\F} f^{1/8+\F} l^{-3/8+\F} M_\mathrm{c}^{-9/8-\F}. \label{eq:massspectrum}
\end{eqnarray}
Particularly we find a relation between the distorted number density and that with no accretion as follows:
\begin{equation}
n(M_\mathrm{c}) = \frac{\alpha_\mathrm{i}(M_\mathrm{i})}{\alpha_\mathrm{i}(M_\mathrm{c})} \left(4 M_4^2 \frac{f l}{M_\mathrm{c}}\right)^{\F} n(M_\mathrm{c})|_\text{no accretion} \label{eq:relationofn}.
\end{equation}
We realize that the accretion controls how effectively other braneworld parameters $ l $ and $ f $ modify the mass spectrum of PBHs.

Now we evaluate the parameter-dependent factors in the above expression.
We first recall that variables that have uncertainties on the orders of magnitude are $ \alpha_\mathrm{i} $ and $ l $.
On the other hand $ f $ stays at $ \mathcal{O}(1) $.
Since $ F $ runs from zero to around unity, $ \F $ is typically small: $ 0 \leq \F \leq 9/(8\pi-8) \approx 0.53 $.
Then it appears that for any values of $ \F $, constants and $ f $ in the round brackets in Eq.~(\ref{eq:relationofn}) have little significance on mass spectrum.
Thus we concentrate on $ l $- and $ M_\mathrm{c} $-dependent parts.

At first we shall fix $ l $ and let $ F $ run to investigate the pure effect of accretion.
We see in Eq.~(\ref{eq:massspectrum}) that $ \F $, accretion itself, can modify the spectrum generally softer: from $ \propto \alpha_\mathrm{i}(M_\mathrm{i}) M_\mathrm{c}^{-1.1} $ to maximally $ \propto \alpha_\mathrm{i}(M_\mathrm{i}) M_\mathrm{c}^{-1.7} $.
This reflects the fact that the number of lighter PBHs is more enhanced than that of heavier ones by accretion.
Whether the number density of PBHs with initial mass $ M_\mathrm{c} $ is enhanced or reduced is determined by the proportionality
\begin{equation}
n \propto \frac{\alpha_\mathrm{i}(M_\mathrm{i})}{\alpha_\mathrm{i}(M_\mathrm{c})} \left(\frac{l}{M_\mathrm{c}}\right)^\F \label{eq:numbergrowth}.
\end{equation}
Because typical values of $ l $ and $ M_\mathrm{c} $ which we are now interested are in the ranges $ 1 \ll M_\mathrm{c} \lesssim 10^{20} \lesssim l \lesssim 10^{31} $ in 4D Planck units, the round bracket gives an enhancing factor of orders of magnitude.
On the other hand we have less knowledge of $ \alpha_\mathrm{i} $.
However, we expect $ \alpha_\mathrm{i} $ does not vary so rapidly in $ M $, hence we never get involved in $ \alpha_\mathrm{i} $ too much and hereafter omit its contribution.
Therefore it is concluded that in this case the number of PBHs is actually enhanced and it directly leads to the gain of a resulting diffuse photon spectrum.

Next we fix $ F $ and let $ l $ vary.
When $ l $ varies over orders of magnitude, number density also varies by nearly the same orders as $ l $, $ n \propto l^{\F} $.
However, we should remember that varying $ l $ results in a change of mass range relevant to the observable diffuse photon spectrum.
Thus we derive parameter-dependent behavior of a photon spectrum in the following manner.
First the absolute values of the present photon spectrum depend mainly on the total amount of PBHs having already evaporated away by now because most of the energy emission from individual PBHs occurs around the end of their lifetimes.
Here, since the lifetime $ t_\mathrm{evap} $ is proportional to $ M_\mathrm{c}^2 l $, the initial mass of PBHs with a specific lifetime obeys $ M_\mathrm{c} \propto l^{-1/2} $.
Therefore from Eq.~(\ref{eq:massspectrum}) the number of PBHs which evaporate at some cosmological time obeys $ n \propto \alpha_\mathrm{i}(M_\mathrm{i}) l^{3/16+3\F/2} $.
It is noted here that the absolute value of the resulting photon spectrum is related not only to the number density of evaporating PBHs but to their mass and temperature.
Now we shall concentrate on the peak value of the spectrum and investigate its $ l $ and $ F $ dependences.
In order to evaluate the peak, we here make an approximation that photons relevant to the high-energy peak are only those emitted from PBHs just evaporating now.\footnote{
The intensity of the emitted particle shows a peak at $ \sim 5 \times T_\mathrm{BH} $ in the 4D case \cite{Page1976}.
}
Furthermore, we also regard that the number of emitted photons per PBH is proportional to $ M_\mathrm{c}/T_\mathrm{BH} $.
Thus the dependence of the peak value $ I_\mathrm{peak} $ on parameters is obtained as
\begin{eqnarray}
I_\mathrm{peak}
 &\propto& n(M_\mathrm{c}^*) \frac{M_\mathrm{c}^*}{T_\mathrm{BH}^*} \nonumber \\
 &\propto& \alpha_\mathrm{i}(M_\mathrm{i}^*) l^{-1/16+3\F/2} \label{eq:peakvalue}.
\end{eqnarray}

We further derive the relation between $ I_\mathrm{peak} $ and Hawking temperature $ T_\mathrm{BH}^* $, which describes a trajectory of peaks when $ l $ is varied but $ F $ is fixed.
This representation will turn out to be greatly useful in later analyses.
Substituting $ l \propto T_\mathrm{BH}^{*~-4} $, the resulting relation is given as
\begin{eqnarray}
I_\mathrm{peak}
 &\propto& \alpha_\mathrm{i}(M_\mathrm{i}^*) T_\mathrm{BH}^{*~1/4-6\F} \nonumber \\
 &\equiv& \alpha_\mathrm{i}(M_\mathrm{i}^*) E_0^p, \label{eq:trajectory}
\end{eqnarray}
where we changed the variable from $ T_\mathrm{BH}^* $ to $ E_0 $, and defined $ p \equiv 1/4 - 6\F $.
The region for the parameter $ E_0 $ is limited to $ 10~\text{keV} \lesssim E_0 \lesssim 100~\text{MeV} $ as in Eq.~(\ref{eq:Hawkingtemperature^*region}).
We find that there is a specific value $ F \approx 0.11 $ that sets the above index to be zero.
However, in our framework, it will appear that a more significant value is $ F \approx 0.52 $ giving the index $ p \approx -1.1 $, which is actually the index of the power-law fitting of the diffuse $ \gamma $-ray background observed by EGRET \cite{Sreekumaretal1998}.

\subsection{Comparing with observations}

From now on, we compare the predicted spectra with the observed diffuse photon background and then obtain the upper limits on the mass fraction $ \alpha_\mathrm{i} $ by requiring that those spectra do not contradict the observation.
$ \mathrm{LIM_i} $s are determined for each parameter set.

First of all, the observed high-energy diffuse photon background spectrum in the region $ 100~\text{keV} $ - $ 1~\text{GeV} $ obeys approximately the power-law with index $ -1 $ to $ -2 $ \cite{Gruber1992,Kinzeretal1997,Kappadath1998,Sreekumaretal1998}.
On the other hand it was shown by Clancy {\it et al.} \cite{CGL2003} that a photon spectrum emitted by PBHs asymptotically approaches $ \sim E_0^{-3} $ in the high energy tail, and has typically more hard shape than $ E_0^{-1} $ in the energy range lower than the peak as long as $ \alpha_\mathrm{i} $ does not vary so quickly with $ M_\mathrm{i} $.
Therefore under the same assumption it is understood that the spectral peak generally gives an upper bound on $ \alpha_\mathrm{i} $.
Keeping this fact in mind, we further simplify so that $ \alpha_\mathrm{i} $ apparently does not depend on $ M_\mathrm{i} $ in the following analysis.

\begin{figure}[ht]
\begin{center}
\includegraphics[width=8cm]{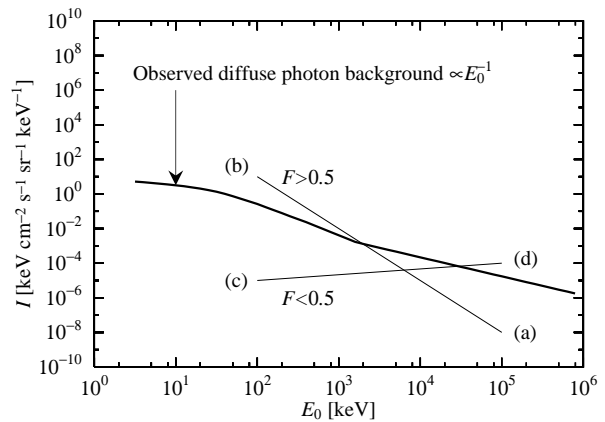}
\end{center}
\caption{\label{fig:pon}
A schematic view of the relations between the diffuse photon background and two kinds of trajectories.
The peak energy is limited in the region shown in Eq.~(\ref{eq:Hawkingtemperature^*region}), thus the peak can move between either pair of ends.
In each case $ \alpha_\mathrm{i} $'s are set to be large enough to make trajectories and the photon background cross.
}
\end{figure}

In this case the peak trajectory Eq.~(\ref{eq:trajectory}) is fixed except with respect to the ``normalization factor'' $ \alpha_\mathrm{i} $.
Here, consider the case in which the trajectory and the observed background spectrum have an intersection at some large $ \alpha_\mathrm{i} $.
Apparently the $ l $ corresponding with the intersection bounds the size of the extra dimension.
At this point we realize that it is possible to classify the situations into two distinctive classes; one is such that the gradient of the peak trajectory $ p \lesssim -1 $, i.e., $ F \gtrsim 0.5 $, and the other is $ p \gtrsim -1 $, $ F \lesssim 0.5 $.
The scheme is shown in Fig.~\ref{fig:pon}.

In the former case (``$ F > 0.5 $''), the peak trajectory is ``softer'' than that of the observed background spectrum.
We should recall that higher energy corresponds with smaller $ l $.
Thus once the value of $ l $ at the intersection is given, it sets an {\it upper bound} on $ l $; the size of the extra dimension $ l $ is allowed only if it is smaller than this bound so that the peak is hidden in the photon background.
Saying with the aid of Fig.~\ref{fig:pon}, the peak locates in the side of (a) rather than (b).
On the other hand, in the latter case (``$ F < 0.5 $''), the trajectory is ``harder'' than the background.
It can be even an increasing function of $ E_0 $.
Somewhat surprisingly, this time the intersection gives a {\it lower bound} on $ l $.
Namely, a rather large extra dimension is required, i.e., (c) is allowed but (d) is rejected.
However we again should note that the above discussions are valid only when the trajectory intersects the background.
Since $ E_0 $ moves in a limited range, a much smaller $ \alpha_\mathrm{i} $ prevents them from crossing.

For further quantitative investigation, we consider the {\it transition} between five dimensions and four dimensions.
To begin with, we determine when and how four-dimensionality is recovered in our framework in the following manner.
Although we naturally expect that everything should recover four-dimensionality in the small $ l $ limit, when the recovery does occur is not trivial because it depends on what quantity we treat.
In our framework, the observable quantity is the shape of the photon spectrum emitted from PBHs in the past.
There are especially two elements, the absolute value and the location of the peak.
The former depends on both the size of the extra dimension $ l $ and the accretion efficiency $ F $, while the latter is determined by $ l $ alone.
The peak energy leaves the 4D value only when the Hawking temperature of PBHs evaporating now differs from the 4D one.
The difference in Hawking temperature comes from the inequality of the Schwarzschild radius $ r_\mathrm{S} $ of the PBH vs the AdS curvature radius $ l $.
If a PBH has sufficiently smaller $ r_\mathrm{S} $ than $ l $, it is well described as a 5D object and its Hawking temperature deviates from the 4D case according to Eqs.~(\ref{eq:Hawkingtemperature}) or (\ref{eq:Hawkingtemperature^*}).
Therefore the change of peak location can occur when PBHs evaporating now have radii smaller than $ l $.
However, in the opposite case where they have a larger radius than $ l $, their temperature and the evolution are the same as in the 4D case.
Although a spectral shape other than the peak can be modified a little by the 5D effect, properties of the peak, such as the location and its absolute value, are indistinguishable from 4D ones.
Thus we can set the boundary between 5D and 4D on the case when the size of the Schwarzschild radius of newly evaporating PBHs agree with the size of the extra dimension: $ r_\mathrm{S}=l $, which corresponds to $ l \approx 5.1 \times 10^{19} l_4 $.
With an additional fact that PBHs formed at $ t \sim t_\mathrm{c} $ have $ r_\mathrm{S} \sim l $, this condition is equivalent to the one that the 5D accretion effect becomes significant, as shown in Sec.~\ref{sec:formulation}.

By the way we have already found that there is some region in parameter space with low accretion efficiency ($ F \lesssim 0.5 $) in which a smaller extra dimension is disfavored.
Of course the existence of such a strange region is entirely due to the existence of the extra dimension, but from the above arguments, it was revealed that there is some critical size of $ l $ in our framework below which the effect of the extra dimension is hidden.
Therefore we conclude that the {\it favorable} $ l $ with low accretion efficiency is at least larger than the critical size $ 5.1 \times 10^{19} l_4 $.

We can confirm the above implications by taking upper bounds on $ \alpha_\mathrm{i} $ into account specifically.
Let us first consider the case of $ F \lesssim 0.5 $, in which the peak trajectory obeys $ I_\mathrm{peak} \propto E_0^p $ with the index $ p > -1 $.
The severest bound on $ \alpha_\mathrm{i} $ is set when the peak locates at the high energy end of its trajectory, i.e., $ \sim 100~\text{MeV} $, and comes in touch with the observed background spectrum there.
Then the effective 4D condition should hold and the corresponding upper bound on $ \alpha_\mathrm{i} $ is exactly the same as that of the 4D case, $ 10^{-27} $ \cite{GreenLiddle1997}.
On the other hand, if the Universe has an extra dimension with a size exceeding the critical value, the peak moves to a lower energy region sufficiently hidden in the background spectrum and hence an $ \alpha_\mathrm{i} $ larger than $ 10^{-27} $ becomes allowed.
As a consequence $ \alpha_\mathrm{i} $ larger than $ 10^{-27} $ requires $ l $ to be larger than $ 5.1 \times 10^{19} l_4 $.
Note that the above conclusion makes sense only when the conditions $ \alpha_\mathrm{i} \gtrsim 10^{-27} $ and $ F \lesssim 0.5 $ are simultaneously realized.
The former is forbidden in the pure 4D universe but the latter means that the PBH mass spectrum is almost the same as that of the 4D case, where no accretion works.
Therefore the existence of a large extra dimension, which can decelerate the evolution of PBH, is needed to avoid conflict.
On the contrary no lower bound on $ l $ is set when $ \alpha_\mathrm{i} $ is smaller than $ 10^{-27} $.

Finally, we mention the case $ F\gtrsim 0.5 $.
Like the $ F \lesssim 0.5 $ case, the peak trajectory intersects the observed background spectrum at the highest energy end when $ \alpha_\mathrm{i}=10^{-27} $.
However, this time the trajectory entirely locates above the background spectrum except at the intersection.
Thus in order to make the extra dimension larger than $ 5.1 \times 10^{19} l_4 $ and let the peak move to a lower energy region ensuring observed results, the trajectory must fall down by setting $ \alpha_\mathrm{i} $ lower than $ 10^{-27} $.
Consequently $ \alpha_\mathrm{i} \gtrsim 10^{-27} $ is never allowed and $ l $ larger than $ 5.1 \times 10^{19} l_4 $ gives a rather stringent upper bound on $ \alpha_\mathrm{i} $ than in the 4D case.
Of course there is no deviation from the 4D case if $ l $ is below the critical size.

\section{\label{sec:numerical}
NUMERICAL RESULTS
}

Here we confirm and refine the arguments in the previous section by virtue of numerical calculation.
Relevant formulas are Eqs.~(\ref{eq:U_0}) and (\ref{eq:I}) with Eq.~(\ref{eq:massspectrum}).
At first, we study how the spectral shape of the PBH photon depends on each kind of parameter, and then observationally allowed regions for braneworld parameters and their implications are shown.

In the following calculations, we assume that $ \alpha_\mathrm{i} $ is constant in $ M_\mathrm{i} $ for simplicity.
We finish calculating the evaporation process of the individual PBH when its Hawking temperature reaches $ m_\pi \approx 135~\text{MeV} $.

Observational data being used in this section are listed in Table~\ref{tab:Obs}.
We should note that there is relatively large error in the range $ 400~\text{keV--} 30~\text{MeV} $ among these observations.
There remains uncertainty in our results coming from it.

\begin{table}[ht]
\caption{\label{tab:Obs}
Observational data used for patches of energy regions and their references.
}
\begin{ruledtabular}
\begin{tabular}{rcrc}
$      3~\mathrm{keV} $ &-& $     60~\mathrm{keV} $ & HEAO-1 A2 \cite{Gruber1992} \\
$     80~\mathrm{keV}\footnotemark $
                        &-& $    450~\mathrm{keV} $ & HEAO-1 A4 \cite{Kinzeretal1997} \\
$    450~\mathrm{keV} $ &-& $ \sim 1~\mathrm{MeV} $
                                                  & SMM\footnotemark \cite{Kappadath1998} \\
$ \sim 1~\mathrm{MeV} $ &-& ~                     & EGRET \cite{Sreekumaretal1998} \\
\end{tabular}
\end{ruledtabular}
\footnotetext[1]{There seems to be no appropriate data in $ 60\text{--}80~\text{keV} $, but this energy range is not a concern in our discussions.
We just draw the interpolated spectrum in this region.}
\footnotetext[2]{Since there are relatively large uncertainties in this energy region, we use only the index of SMM and connect the values of HEAO-1 A4 and EGRET.
This method may cause an error of an order of magnitude in our conclusion.
Note that such a spectrum is consistent with COMPTEL \cite{Kappadath1998}.}
\end{table}

\subsection{Accretion efficiency $ F $}

\begin{figure}[ht]
\begin{center}
\includegraphics[width=8cm]{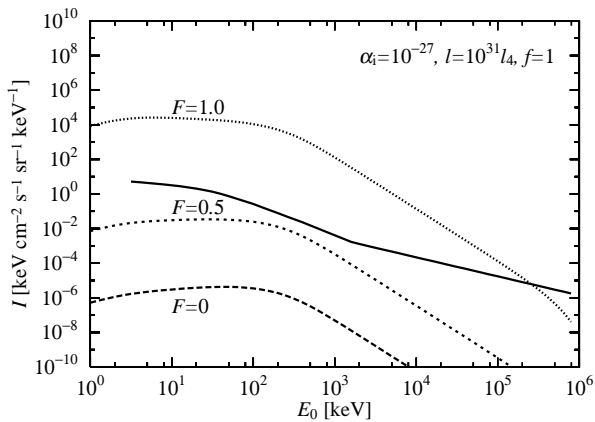}
\end{center}
\caption{\label{fig:I_F}
Spectra of surface brightness $ I $ for $ F=0 $ (long-dashed), $ 0.5 $ (short-dashed), and $ 1.0 $ (dotted), respectively.
Other parameters are fixed: $ \alpha_\mathrm{i} = 10^{-27} $, $ l = 10^{31} l_4 $, and $ f=1 $.
The solid line indicates an observed diffuse photon background.
Spectra with accretion ($ F=1.0 $ and $ 0.5 $) are almost uniformly gained according to Eq.~(\ref{eq:numbergrowth}) relative to that with no accretion ($ F=0 $).
A slight difference of gradients in the low energy region between these spectra is seen; it comes from the fact that the number of lighter mass PBHs is relatively more enhanced by accretion than that of heavier PBHs.
}
\end{figure}

We first discuss the effect of accretion on PBHs in the brane high energy phase.
Changes in the spectra caused by varying $ F $ are shown in Fig.~\ref{fig:I_F}.\footnote{
Note that hereafter we set the value of the initial mass fraction $ \alpha_\mathrm{i} $ to be $ 10^{-27} $.
Those spectra crossing the observed diffuse photon background are observationally rejected for this {\it standard} $ \alpha_\mathrm{i} $, e.g., the $ F=1.0 $ case in Fig.~\ref{fig:I_F}.
}
Accretion efficiency $ F $ is set to $ 0 $, $ 0.5 $, and $ 1.0 $, while other parameters are fixed to $ \alpha_\mathrm{i}=10^{-27} $, $ l=10^{31} l_4 $, and $ f=1 $.
Because other parameters $ l $ and $ f $ are fixed, the initial PBH mass spectrum and subsequent evolution of individual PBHs are both common.
In addition, the mass range relevant to the resulting diffuse photon spectrum, namely of PBHs with lifetime $ t_\mathrm{dec} \leq t_\mathrm{evap} \leq t_0 $, is never changed.

The spectra show two peaks as mentioned in Sec.~\ref{sec:formulation}.
The high-energy peak is around $ (\text{a few}) \times 100~\text{keV} $, which corresponds to the Hawking temperature of a PBH with lifetime $ t_0 $.
See Eq.~(\ref{eq:Hawkingtemperature^*}) and substitute $ l = 10^{31} l_4 $.
The enhancement of absolute value in this case comes from the number growth of PBHs as in Eq.~(\ref{eq:numbergrowth}).
Specifically, making use of Eq.~(\ref{eq:M_c^*}) with $ l = 10^{31} l_4 $, we obtain factors of
\begin{subequations}
\begin{eqnarray}
\left(\frac{l}{M_\mathrm{c}}\right)^\F
 &=& 10^{8.9} ~~~ \text{for}~F=1.0 \\
 &=& 10^{3.6} ~~~ \text{for}~F=0.5,
\end{eqnarray}
\end{subequations}
which explain calculated spectra well.

The change of gradient in the lower energy region is due to the distortion of the initial mass spectrum.
In short because of relatively large enhancement of smaller PBHs.

\subsection{Bulk AdS curvature radius $ l $}

Next we consider the dependence on bulk AdS curvature radius $ l $, one of the most fundamental quantities in the braneworld scenario.

At first glance, one finds two basic facts, that (i) the Hawking temperature of a black hole on the brane is lowered as $ \propto l^{-1/4} $ in contrast to the 4D case, and that (ii) the brane high energy phase ($ t \leq l/2 $) becomes longer for larger $ l $.
Thus in general the behavior is rather complicated.
However, we can neglect the second fact when accretion does not work.
Hence we first concentrate on cases in which accretion never works ($ F = 0 $), then we consider the composite cases ($ F = 1.0~\text{and}~0.5 $) later.

\begin{figure}[ht]
\begin{center}
\includegraphics[width=8cm]{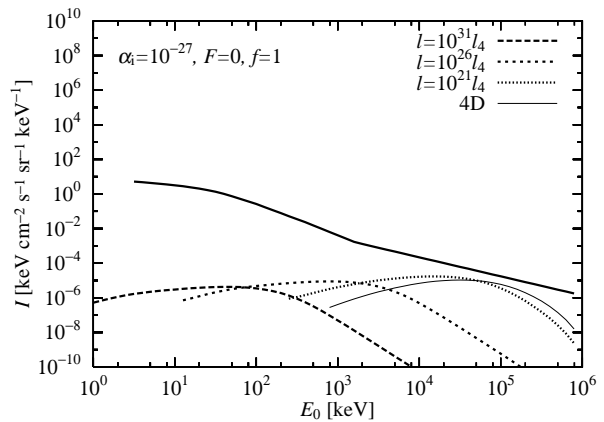}
\end{center}
\caption{\label{fig:I_l_0}
Spectra of surface brightness $ I $ for $ l=10^{31} l_4 $ (long-dashed), $ 10^{26} l_4 $ (short-dashed), and $ 10^{21} l_4 $ (dotted), respectively.
The other parameters are fixed: $ \alpha_\mathrm{i}=10^{-27} $, $ F=0 $, and $ f=1 $.
A 4D spectrum (thinsolid line) is also shown.
The high-energy spectral peaks move from high energy (right) to low energy (left) as the extra dimension becomes large.
Actually, the peaks draw the trajectory $ I_\mathrm{peak} \propto E_0^{0.25} $ according to Eq.~(\ref{eq:trajectory}).
As seen in this figure, it is expected that the upper bound on $ \alpha_\mathrm{i} $ for this parameter set of $ F $ and $ f $ is given in the high energy region, and thus in turn large $ l $ is favored.
Convergence on the 4D spectrum as $ l $ approaches $ 10^{20} l_4 $ is also shown.
}
\end{figure}

Figure \ref{fig:I_l_0} shows dependence of spectral shape on AdS curvature radius $ l $ in the case of no accretion.
$ l $ is taken to be $ 10^{31} l_4 $, $ 10^{26} l_4 $, and $ 10^{21} l_4 $; $ f $ and $ F $ are fixed to unity and zero, respectively.
The 4D spectrum is also shown in the figure.

At first we restrict ourselves to the three 5D spectra.
Since there is no accretion, mass spectra of PBHs in each case are identical.
They are expressed as Eq.~(\ref{eq:massspectrum}) with $ \F=0 $.
The only difference between the three spectra comes from the Hawking temperature, equivalently the lifetime of PBH.
For larger $ l $, the Hawking temperature of PBHs is lower.
Thus the number of PBHs contributing to the peak is enhanced, but on the other hand the number of photons emitted from individual PBHs is decreased.
The two opposite effects almost compensate in this case, see Eq.~(\ref{eq:peakvalue}) and set $ \F=0 $.
As a consequence, the peaks draw an almost horizontal trajectory $ I_\mathrm{peak} \propto E_0^{0.25} $ according to Eq.~(\ref{eq:trajectory}).

Next we pay attention to the 4D spectrum.
We can confirm the consistency of our calculation with the 4D case in Fig.~\ref{fig:I_l_0} from the converging behavior of spectra upon the 4D spectrum as $ l $ becomes small.
As mentioned in the previous section, the transition from 5D to 4D occurs when the 5D approximation for the PBHs with lifetime $ t_0 $ breaks down.
The critical size of $ l $ is $ \approx 5.1 \times 10^{19} l_4 $.
The properties of PBHs with lifetime $ t_\mathrm{evap}=t_0 $ under such a condition,
\begin{equation}
T_\mathrm{BH}^* = 38~\text{MeV} ~~~ \text{and} ~~~ M_\mathrm{c}^* = 1.3 \times 10^{15}~\text{g},
\end{equation}
are both consistent with known 4D values.
In addition, under this 4D limit, the brane high energy phase is shortened and number growth of PBHs hardly occurs.
As a consequence 5D spectra succeedingly reproduce the shape of 4D spectrum with respect to both the location of its peak and absolute value under the 4D limit.

Note that the 4D spectrum sets the upper limit on the mass fraction $ \alpha_\mathrm{i} $ around $ \sim 10^{-27} $ \cite{GreenLiddle1997}.
We already argued in the previous section that the value $ 10^{-27} $ takes a crucial role even in the braneworld.
We later confirm it numerically.

\begin{figure}[hb]
\begin{center}
\begin{tabular}{c}
\includegraphics[width=8cm]{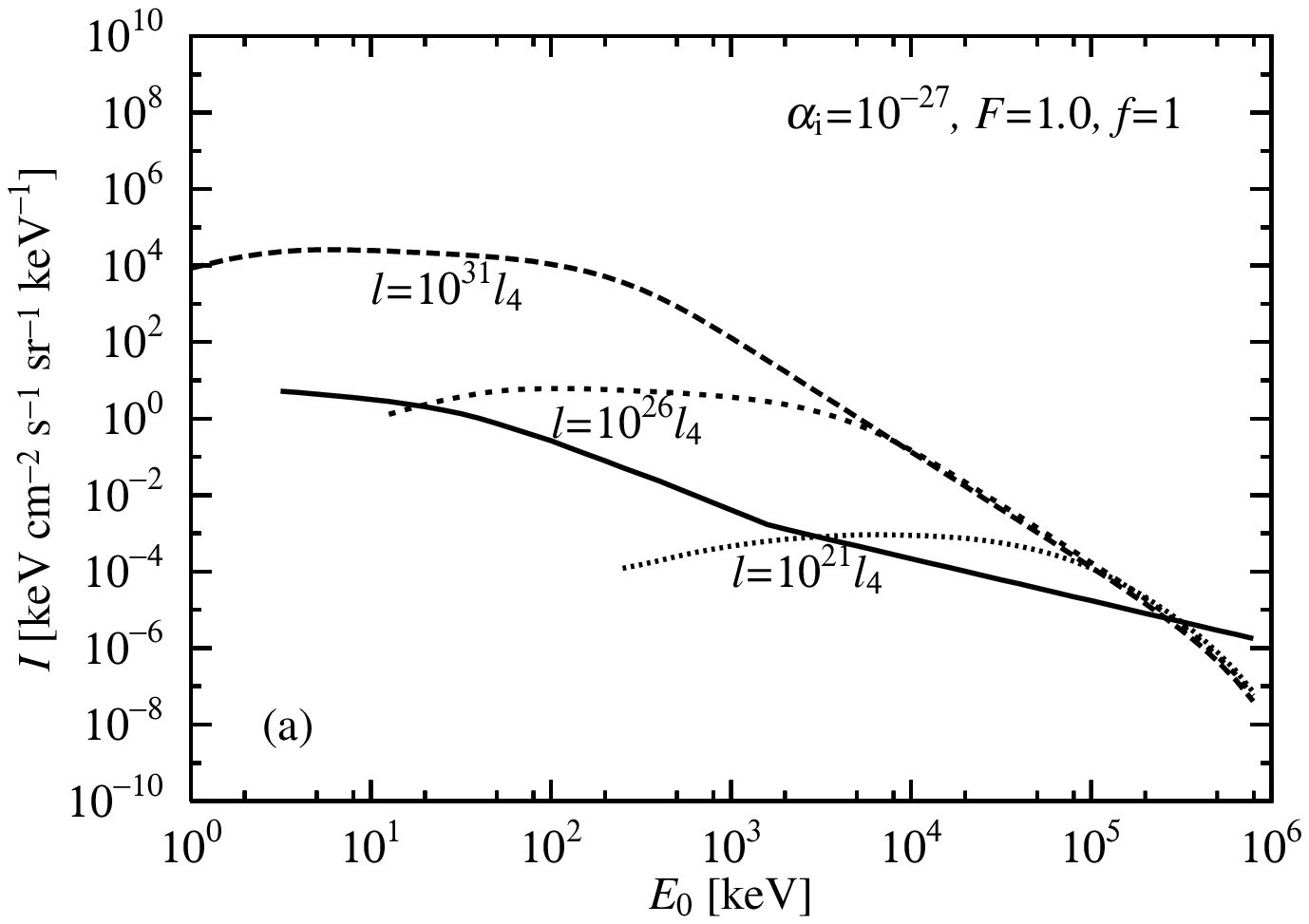} \\
\includegraphics[width=8cm]{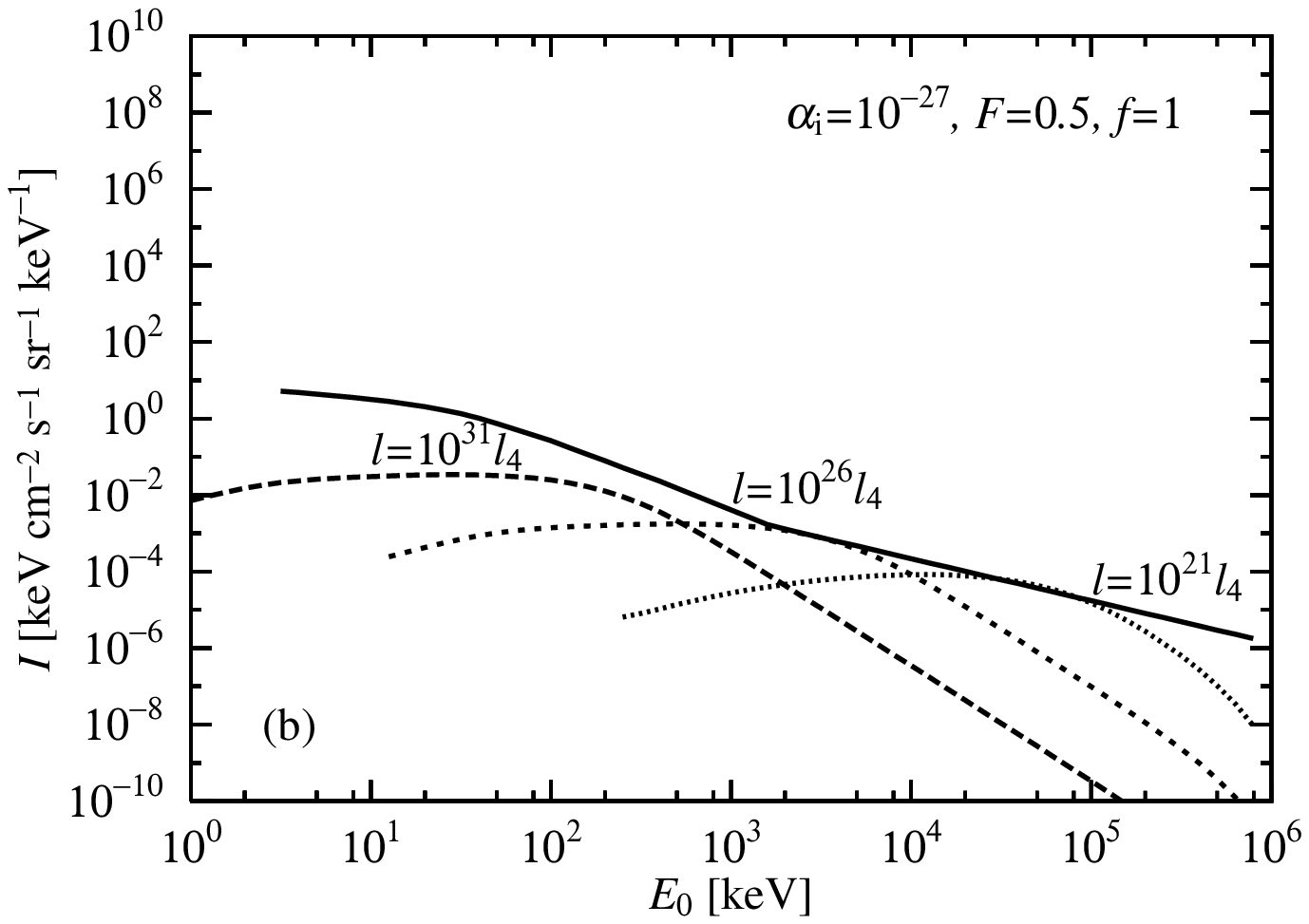}
\end{tabular}
\end{center}
\caption{\label{fig:I_l_1}
Spectra of surface brightness $ I $ for $ l=10^{31} l_4 $ (long-dashed), $ 10^{26} l_4 $ (short-dashed), and $ 10^{21} l_4 $ (dotted), respectively, where the other parameters are fixed: $ \alpha_\mathrm{i}=10^{-27} $, $ f=1 $, (a) $ F=1.0 $, and (b) $ 0.5 $.
In both figures, the spectral peaks or edges move from the high energy region (right) to the low energy region (left) as the extra dimension becomes large.
The peaks in two figures draw trajectories $ I_\mathrm{peak} \propto E_0^{-2.9} $ [(a) $ F=1.0 $] and $ E_0^{-1.0} $ [(b) $ F=0.5 $] according to Eq.~(\ref{eq:trajectory}).
By considering the gradients of trajectories, we find that upper bounds on $ l $ are given for $ F=1.0 $ from the value of $ l $ corresponding with the possible intersection of the observation and trajectory; peaks must locate in the right of the intersection.
In particular, $ \alpha_\mathrm{i} < 10^{-27} $ is required to realize $ l > 5.1 \times 10^{19} l_4 $.
Note that in the $ F=0.5 $ case, the gradient of the peak trajectory and that of the observed spectrum are almost the same.
}
\end{figure}

By the way, now we are ready to understand the general behavior of spectral shape for varying $ l $ together with the existence of accretion.
See Fig.~\ref{fig:I_l_1}.

The locations of peak energy are explained in the same manner as in the case of no accretion.
In addition, the dependence of absolute values on parameter sets is explained by Eq.~(\ref{eq:peakvalue}) as
\begin{subequations}
\begin{eqnarray}
I_\mathrm{peak}
 &=& l^{0.72} ~~~ \text{for}~F = 1.0\\
 &=& l^{0.26} ~~~ \text{for}~F = 0.5,
\end{eqnarray}
\end{subequations}
and the trajectories of their peaks are
\begin{subequations}
\begin{eqnarray}
I_\mathrm{peak}
 &=& E_0^{-2.9} ~~~ \text{for}~F = 1.0\\
 &=& E_0^{-1.0} ~~~ \text{for}~F = 0.5.
\end{eqnarray}
\end{subequations}

We can see that in the case $ F=1.0 $ an {\it upper} bound on $ l $ is given at the intersection of the above trajectory and the observed background spectrum.
To the contrary, in the case $ F=0 $, a {\it lower} bound is obtained from the intersection.
On the other hand the spectral index for $ F=0.5 $ is nearly identical to that of the observed background.
Thus we confirm the mention in the previous section that the bound on $ l $ switches its character as an upper bound to that as a lower bound when accretion efficiency falls down across 50\%.

\subsection{Uncertainty of formation process}

The horizon mass fraction parameter $ f $ takes a less important role in the results.
It only contributes as an overall factor for the number density of PBHs in the form $ n \propto f^{1/8+\F} $ as shown in Eq.~(\ref{eq:massspectrum}).
From a physical point of view, $ f $ has a parameter range at most within a few orders of magnitude.
Therefore we skip showing the spectral shapes for varying $ f $; they are not so interesting.

\subsection{Allowed regions: The final result}

As shown in this section, we have obtained photon spectra for sufficiently many parameter sets.
Here we apply those {\it sample} spectra to determine how large the initial mass fraction $ \alpha_\mathrm{i} $ can be.

\begin{figure}[ht]
\begin{center}
\includegraphics[width=8cm]{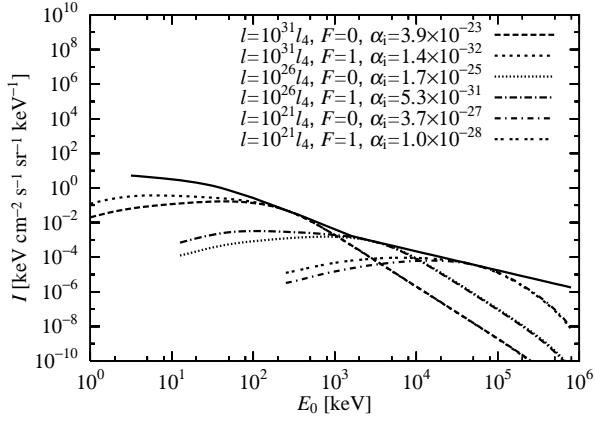}
\end{center}
\caption{\label{fig:adjusted}
Illustration of how to obtain $ \mathrm{LIM_i} $'s.
The six spectra in this figure are those that appeared in Fig.~\ref{fig:I_l_0} and Fig.~\ref{fig:I_l_1}(a), the $ F=1.0 $ case, but this time their absolute values are adjusted to be in contact with the observed spectrum, i.e., $ \alpha_\mathrm{i} \neq 10^{-27} $.
These $ \alpha_\mathrm{i} $'s give upper bounds on the fraction of PBHs for each parameter set.
}
\end{figure}

To achieve it, we adjust the calculated spectra to come in contact with the observational data, then we can determine the value of the upper limit on $ \alpha_\mathrm{i} $, $ \mathrm{LIM_i} $, by measuring overall factors of adjusted spectra.
The method is shown in Fig.~\ref{fig:adjusted}.

It is confirmed in Fig.~\ref{fig:adjusted} that the most crucial part of each spectrum giving the upper limit on $ \alpha_\mathrm{i} $ is nothing but the high-energy peak.

\begin{figure}[hbt]
\begin{center}
\includegraphics[width=8cm]{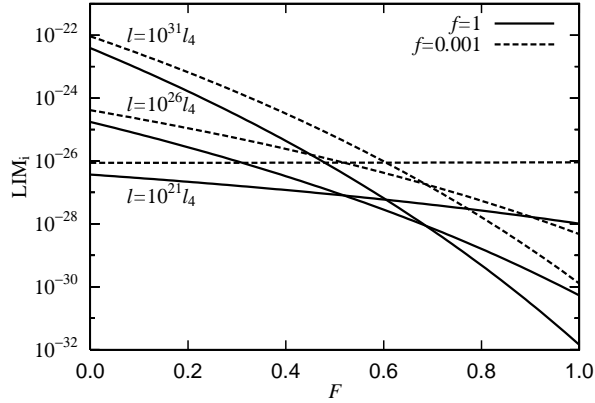}
\end{center}
\caption{\label{fig:LIM_i}
Illustration of how $ \mathrm{LIM_i} $'s depend on $ F $'s in the two cases $ f=1 $ (solid) and $ f=0.001 $ (dashed).
For each $ f $, three sizes of extra dimension are taken: $ l=10^{31} l_4 $, $ 10^{26} l_4 $, and $ 10^{21} l_4 $ from above to below in the figure.
The differences of qualitative behavior between the lines are coming from the sensitiveness to accretion efficiency.
For a larger extra dimension, the duration of the brane high energy phase is lengthened and hence the effect of accretion is stressed.
To the contrary, as the size of the extra dimension becomes small accretion ceases to work for any values of $ F $, and $ \mathrm{LIM_i} $ becomes less sensitive to $ F $.
$ \mathrm{LIM_i} $ is almost independent of $ F $ under the limit $ l \rightarrow 10^{20} l_4 $.
}
\end{figure}

Collecting values of $ \mathrm{LIM_i} $ for $ F $'s, we can show the $ F $-dependence of $ \mathrm{LIM_i} $ as in Fig.~\ref{fig:LIM_i}, where we selected two $ f $'s, 1 and 0.001.
In this figure we find that $ \mathrm{LIM_i} $'s fall as $ F $'s become large for almost all the parameter sets displayed here, while one set $ l=10^{21} l_4 $ and $ f=0.001 $ shows subtle behavior.
It should come from a detailed spectral shape around the peak, thus beyond our approximated analysis.
When accretion in the brane high energy phase works efficiently, the number of PBHs grows along with increasing $ F $, and it leads directly to a gained peak.
Therefore the upper limit on $ \alpha_\mathrm{i} $ is lowered in order to match observations.
On the other hand as $ l $ becomes small, accretion does not have an influence on the resulting spectra and then $ \mathrm{LIM_i} $ becomes independent of $ F $.
Again we confirm here the importance of 50\% accretion efficiency in Fig.~\ref{fig:LIM_i}, which can be read from the behaviors of lines; the lines all pass through around the point $ F=0.5 $ and $ \mathrm{LIM_i}=10^{-27} $.

The constraint $ \mathrm{LIM_i} = 10^{-23} $ obtained by Clancy {\it et al.} \cite{CGL2003} corresponds with the point $ (F=0, \mathrm{LIM_i} \approx 3.9 \times 10^{-23}) $ on a line with $ l=10^{31} l_4 $ and $ f=1 $.
Apparently the constraint is relaxed by some numerical factor for this case.

\begin{figure}[htb]
\begin{center}
\includegraphics[width=8cm]{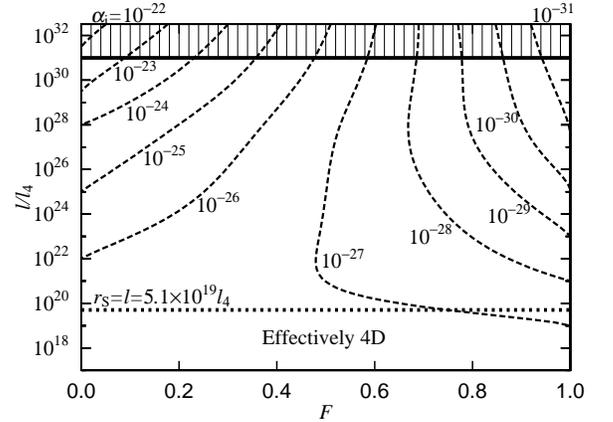}
\end{center}
\caption{\label{fig:region}
Showing of boundaries for $ \alpha_\mathrm{i}=10^{-22}\text{--}10^{-31} $ from left to right.
The remaining parameter $ f $ is here fixed to unity.
The critical size of extra dimension at which transition from 5D to 4D occurs is also shown (dotted line).
The experimentally excluded region $ l \geq 10^{31} l_4 $ is shaded.
The boundaries split the parameter plane into two regions, respectively, the left side is allowed and the right side is excluded.
Once PBHs' mass fraction $ \alpha_\mathrm{i} $ is determined in some way, this diagram indicates what region in the space of braneworld parameters is allowed.
Further information obtainable from this diagram is argued in the text.
}
\end{figure}

At last, we obtain the final result.
Until now we have taken a stance on searching for upper limits on the mass fraction of PBHs.
However, once the actual value of $ \alpha_\mathrm{i} $ is obtained or constrained in some way, one can determine whether some parameter combination $ (l,F,f) $ is allowed or not by comparing it with $ \mathrm{LIM_i} $ of the set.
Namely the combination will be observationally rejected if $ \alpha_\mathrm{i} > \mathrm{LIM_i} $.
Thus we obtain a tool for limiting the braneworld parameters through the future knowledge on $ \alpha_\mathrm{i} $.
We indicate the diagram with boundaries for ten $ \alpha_\mathrm{i} $ cases in Fig.~\ref{fig:region}.
Each boundary separates the parameter space into two regions; the left side is allowed and the opposite is excluded.

Let us investigate this diagram further.
To begin with, we should confirm the consistency of this diagram with the 4D case, in which $ \mathrm{LIM_i} \sim 10^{-27} $ was obtained by Green and Liddle \cite{GreenLiddle1997}.
As was confirmed in Fig.~\ref{fig:LIM_i}, we expect that our results under the 4D limit should be independent of $ F $.
Actually we see such a signature in Fig.~\ref{fig:region} from the fact that all boundaries go to either the $ F=0 $ axis or $ F=1 $ before they reach $ l=5.1 \times 10^{19} l_4 $.
The above qualitative behaviors denote that under the 4D limit $ l \lesssim 5.1 \times 10^{19} l_4 $ all the sets of braneworld parameters with $ \alpha_\mathrm{i} \gtrsim 10^{-26} $ are rejected, while with $ \alpha_\mathrm{i} \lesssim 10^{-27} $ are all allowed.
Therefore we find that our result under the 4D limit is well consistent with the pure 4D result $ \mathrm{LIM_i} \sim 10^{-27} $.

Furthermore, there is a critical boundary between $ \alpha_\mathrm{i}=10^{-26} $ and $ 10^{-27} $, which separates the whole region into two distinctive patches.
Note that we can identify this critical boundary with a vertical line $ F \approx 0.5 $.
In the left side of the critical boundary, that is $ \alpha_\mathrm{i} \geq 10^{-26} $, all boundaries hit $ F=0 $, while for $ \alpha_\mathrm{i} \leq 10^{-27} $, all hit $ F=1 $.
Therefore we find that only the region $ l \gtrsim 5.1 \times 10^{19} l_4 $, $ F < 0.5 $ is allowed for $ \alpha_\mathrm{i} $'s higher than the critical value.
It means that if large $ \alpha_\mathrm{i} $, which is not allowed in the 4D case, is actually realized, large extra dimension is {\it required} and $ l $ has a lower bound $ \gtrsim 5.1 \times 10^{19} l_4 $.
On the other hand the boundary with $ \alpha_\mathrm{i}=10^{-27} $ asymptotically approaches to $ l=5.1 \times 10^{19} l_4 $.
This means that an extra dimension with $ l $ larger than $ 5.1 \times 10^{19} l_4 $ requires $ \alpha_\mathrm{i} $ to be smaller than $ 10^{-27} $.
Hence a large extra dimension gives more stringent upper bound on $ \alpha_\mathrm{i} $ than the 4D case.
Of course modestly large $ \alpha_\mathrm{i} $ in turn gives a rather severe bound on the size of extra dimension.
For example, if $ \alpha_\mathrm{i}=10^{-27} $ with $ F=1.0 $ is realized, $ l $ is bounded below $ 10^{20} l_4 $, which is about 11 orders of magnitude severer than the present bound obtained from measurement of the short-range gravitational force.
However, one should always keep in mind that if there is merely a smaller initial PBH fraction than $ 10^{-27} $, our framework based on the diffuse high-energy photon background has limited significance on the braneworld parameters.
Even in the most efficient accretion case $ F=1.0 $, the small fraction $ \alpha_\mathrm{i}=10^{-31} $ only gives $ l \lesssim 10^{30} l_4 $, which is almost identical with the known upper bound.
Moreover, $ F < 0.5 $ gives nothing new if $ \alpha_\mathrm{i} \lesssim 10^{-27} $.

\section{\label{sec:conclusion}
CONCLUSIONS AND DISCUSSIONS
}

In this paper, at first we started with the mass spectrum of PBHs at the moments of their formation.
Then we derived the resulting distorted spectrum after accretion in the brane high energy phase.
We found that the PBH number density with specific initial mass $ M_\mathrm{c} $ is enhanced relative to its no-accretion value as
\begin{equation}
n = \frac{\alpha_\mathrm{i}(M_\mathrm{i})}{\alpha_\mathrm{i}(M_\mathrm{c})} \left(4 M_4^2 \frac{f l}{M_\mathrm{c}}\right)^{\F} n|_\mathrm{no~accretion},
\end{equation}
where
\begin{equation}
M_\mathrm{i} = \left(4 M_4^2 f l\right)^{-F/(\pi-F)} M_\mathrm{c}^{\pi/(\pi-F)}.
\end{equation}

Using the result on mass spectrum distortion, we also derived the behavior of the peak value of the diffuse photon spectrum.
Under the assumption that $ \alpha_\mathrm{i}(M_\mathrm{i}) $ is constant, the peak value depends on $ l $, or equivalently on $ T_\mathrm{BH}^{*} $ as
\begin{equation}
I_\mathrm{peak} \propto l^{-1/16+3\F/2} \propto T_\mathrm{BH}^{*~1/4-6\F}.
\end{equation}
The latter proportionality is rather useful when we compare the peak trajectory with the observed power-law diffuse photon background with index $ \approx -1 $.
The exponent of the peak trajectory $ p \equiv 1/4-6\F $ gives a value smaller than $ -1 $ when $ F > 50\% $, while it gives a value larger than $ -1 $ when $ F < 50\% $.
Thus there is a critical accretion efficiency around $ 50\% $, a fairly neutral value.
When $ F $ exceeds $ 50\% $, smaller extra dimension and smaller mass fraction $ \alpha_\mathrm{i} $, typically below the 4D upper bound $ 10^{-27} $, are favored; it is a reasonable situation for the braneworld scenario.
To the contrary, if $ F $ falls below $ 50\% $, large extra dimension is favored.
Furthermore, if actually the mass fraction has a value larger than the 4D upper bound $ 10^{-27} $, the lower bound on the size of extra dimension $ l $ is even set around $ 5.1 \times 10^{19} l_4 $.
This critical value comes from the condition that the Schwarzschild radius of PBHs evaporating now is equal to the size of extra dimension.
We should note that the braneworld signature vanishes regardless of the accretion efficiency provided the extra dimension is smaller than the above critical size.
Naturally in the case $ \alpha_\mathrm{i} \ll 10^{-27} $ the existence of an extra dimension again has nothing to do with our framework.

In this paper we also demonstrated numerical calculations for further preciseness of the above arguments.
In the calculation we achieved mainly two improvements as follows.
First, we used the newest observational cosmological parameters obtained by WMAP, which claims $ \Omega_{\Lambda,0}=0.73 $, $ \Omega_{\mathrm{m},0} h^2=0.135 $, $ h=0.71 $, and $ t_0=13.7~\text{Gyr} $.
Because the observable diffuse photons are emitted by PBHs which evaporated after the decoupling time, $ \Lambda $-domination in the present epoch could affect the resulting spectrum via the accelerated expansion deviated from the Einstein-de Sitter universe (EdS).
However, we found that once a correct assumption on the present age of the Universe $ t_0 \approx 13.7~\text{Gyr} $ is made, the difference of scale factor after the decoupling time is at most only $ (\text{a few}) \times 10\% $ between the two cosmological models.
As a consequence, the deviation of resulting photon spectra was within an order of magnitude.
Second, we precisely took into account the photon emission process and mass decrease of black holes at every moment.
It also could modify the result because the redshift is determined by how long the emitted photons have survived.
In spite of this effect, the resulting spectrum again does not greatly differ from calculations under an approximation of instant evaporation.

After all we obtained numerically precise spectra with an assumption of constant $ \alpha_\mathrm{i} $ in sufficiently wide regions for each braneworld parameter and compared them with relevant observational data.
Some of the calculated spectra were shown in Figs.~\ref{fig:I_F}--\ref{fig:I_l_1}.
The behavior of the peak value and the constraints coming from them were both well explained by the preceding analysis.
We then obtained upper bounds on mass fraction, $ \mathrm{LIM_i} $'s for each parameter set $(l, F, f)$ by the method shown in Fig.~\ref{fig:adjusted}.
$ \mathrm{LIM_i} $ was sketched in Fig.~\ref{fig:LIM_i}.
Our most impressive result is the allowed region in parameter space as shown in Fig.~\ref{fig:region}.
The characters of this diagram were also consistent with our analysis.

Recently some authors argued a possible modification of the Hawking radiation in the context of AdS/conformal field theory (CFT) correspondence in the RS2 braneworld \cite{Emparanetal2002,Tanaka2003,Emparanetal2003}.
They predicted enormously rapid decay of astrophysical black holes due to emission of CFT modes, which is interpreted as the Kaluza-Klein (KK) graviton in the bulk.
However, the applicable range of this effect is rather limited to those PBHs larger than AdS curvature radius $ l $.
Hence, although the transition process of a black hole from 4D to 5D is still not clear, large PBHs are to shed their masses and accumulate in the mass spectrum around $ r_\mathrm{S} \sim l $.
They may enhance the number of PBHs which can contribute to a diffuse photon spectrum and result in a more restrictive upper bound on PBH abundance.
However, by the following estimation it appears that little change is caused in our results.
We here assume that all PBHs larger than $ l $ lose their masses shortly after $ t = t_\mathrm{c} $ and eventually become 5D black holes with $ r_\mathrm{S} \sim M \sim l $, whereas the evaporation process of those PBHs smaller than $ l $ is not affected at all.
This results in the number growth of PBHs of size $ l $, but since in general the realistic number density $ n(M) $ is expected to be steeper than $ M^{-1} $ though there is unknown mass dependence in $ \alpha_\mathrm{i} $, the enhancement of $ n(M)|_{M \sim l} $ is at most of some numerical factor on the order of unity.
Moreover, since PBHs with $ r_\mathrm{S} \sim l $ have lifetime $ t_\mathrm{evap} \sim (l/10^{20}l_4)^3 t_0 $, they have little room to contribute to diffuse photon spectrum with $ l $ in the range we concerned ($ 10^{20} l_4 \lesssim l \lesssim 10^{31} l_4 $).

Although out of our interest in this paper, we point out a possibly meaningful case in the framework of AdS/CFT \cite{Emparanetal2002,Tanaka2003,Emparanetal2003} where $ l \lesssim 10^{20} l_4 $.
Under the same assumptions above, in this case we cannot only observe just evaporating 5D PBHs but also so doing standard 4D ones; they must have entirely decayed before the present epoch.
Therefore it becomes more difficult to detect diffuse photons emitted by PBHs.
Another special case is with critical size of extra dimension, $ l \sim 10^{20} l_4 $.
This time the effect of accumulation around $ M \sim l $ can take a significant role because they should be evaporating just now.
However, at the same time they have a rather high initial temperature $ \sim (\text{a few}) \times 10~\text{MeV} $ and then a fraction of their mass energy poured into photon energy is limited.
It will be necessary to estimate both the compensating two effects.

At last, we mention the simplifications in this paper.
First, we did not take significantly braneworld inflation models into consideration.
Though we sometimes took an assumption that $ \alpha_\mathrm{i} $ is constant in $ M_\mathrm{i} $, the mass spectrum may have a rather complicated shape in reality.
However, if modification is not so radical, it should not affect our argument on braneworld constraints because the most crucial role is taken by those PBHs which evaporate just now.
Second, we ignored the latest time behavior of evaporating black holes.
It may cause an interesting change in spectra, but almost all energy has been already released before that.
Hence it is not so important for the whole discussion above, either.

\begin{acknowledgments}

Y.S. is pleased to acknowledge helpful discussions with Kazuya Koyama.
Y.S. would also like to thank Takeshi Kuwabara and Takashi Hiramatsu for useful discussions.
This work was supported in part through Grant-in-Aid for Scientific Research (S) No.~14102004 and Grant-in-Aid for Scientific Research on Priority Areas No.~14079202 by the Ministry of Education, Science and Culture of Japan.
\end{acknowledgments}


\end{document}